\documentclass[a4paper,12pt]{article}
\usepackage[utf8x]{inputenc} 
\usepackage{jheppub}
\usepackage{amsmath}
\usepackage{slashed}
\usepackage{axodraw}
\usepackage[T1]{fontenc}
\usepackage{comment}
\usepackage{hyperref}
\usepackage{braket}
\usepackage{graphicx}
\usepackage{subfigure}

\title{ A Simultaneous Study of Dark Matter and Phase
Transition: Two-Scalar Scenario}
\author[a]{Karim Ghorbani} {}
\author[b,c]{and  Parsa Hossein Ghorbani}

\affiliation[a]{Physics Department, Faculty of Sciences, Arak University, Arak 38156-8-8349, Iran}
\affiliation[b] {Department of Physics, Faculty of Science, Ferdowsi University of Mashhad, Mashhad, Iran}
\affiliation[c]{Applied Physics Inc., Center for Cosmological Research,
 300 Park Avenue, New York, NY 10022, USA}


\abstract{ The simplest extension of the Standard Model by only one real singlet scalar can explain the observed dark matter relic density while giving simultaneously a strongly first-order electroweak phase transition in the early universe. However, after imposing the invisible Higgs decay constraint from the LHC, the parameter space of the single scalar model shrinks to regions with only a few percent of the DM relic abundance and when adding the direct detection bound, e.g. from XENON100, it gets excluded  completely. In this paper, we extend the Standard Model with two real guage singlet scalars, here $s$ and $s'$, and show that the electroweak symmetry breaking may occur via different channels. Despite very restrictive first-order phase transition conditions for the two-scalar model in comparison to the single scalar model, there is a viable space of parameters in different phase transition channels that simultaneously explains a fraction or the whole dark matter relic density, a strongly first-order electroweak phase transition and still evading the direct detection bounds from the latest LUX/XENON experiments while respecting the invisible Higgs decay width constraint from the LHC.  
}

\begin{document}

\maketitle

\section{Introduction}

There have been numerous attempts in different mainstreams from modifying the theory of gravity to extending the Standard Model (SM) of elementary particles to accommodate the problem of the missing mass or the dark matter (DM). The most successful example of the later is the $\Lambda$CDM model which incorporates the existence of a cosmological constant (responsible for the accelerating expansion of the universe) and a cold dark matter (CDM) as new species of particle(s) living in a dark sector. Within the $\Lambda$CDM model, the weakly interacting massive particle (WIMP) has been specially a successful DM paradigm. A WIMP candidate of dark matter can be embedded in various extensions of the standard model with $\mathbb{Z}_2$ or larger symmetry groups in the hidden (dark) sector. 
In this paper, we investigate the scalar extension of the SM with the $\mathbb{Z}_2$ discrete symmetry group needed for stabilizing the dark matter candidate in the so-called freeze-out mechanism.

The first and the simplest of such models is the extension of the SM with only one real single scalar (only one degree of freedom) which has been studied vastly in the literature, see e.g. \cite{McDonald:1993ex,Burgess:2000yq,McDonald:2001vt, OConnell:2006rsp,Barger:2007im,Profumo:2007wc,Yaguna:2008hd,He:2008qm,Gonderinger:2009jp,Farina:2009ez,Profumo:2010kp,Guo:2010hq,Biswas:2011td,Mambrini:2011ik,Cline:2013gha,Feng:2014vea,Duerr:2015mva,Han:2015hda,Duerr:2015aka,Han:2016gyy, Sage:2016xkb,Cuoco:2016jqt,Casas:2017jjg} in which various phenomenological aspects such as the dark matter relic density extracted from WMAP and Planck \cite{Hinshaw:2012aka,Ade:2013zuv,Aghanim:2018eyx,Athron:2018ipf}, the Higgs invisible decay width from the LHC experiments \cite{Sirunyan:2018owy,Aaboud:2018sfi}, the upper bound on the dark matter elastic scattering cross section off nuclei by XENON100, XENON1T, LUX \cite{Aprile:2012nq,Aprile:2017iyp,Akerib:2013tjd}, gamma rays from annihilation of dark matter interpreted by Fermi-LAT data \cite{Ackermann:2012qk,TheFermi-LAT:2017vmf}, and the theoretical aspects such as perturbativity, vacuum stability, electroweak phase transition, gravitational waves have been investigated. Despite having a small space of parameter, the model is remarkably successful in addressing a subset of the aforementioned constraints. The real scalar field in the single scalar dark matter model plays the role of both the DM candidate and the DM-SM mediator through the Higgs portal.

The status of this model has been  reported by the GAMBIT Collaboration in \cite{Athron:2017kgt}. 
According to the GAMBIT, taking into account all the direct and indirect constraints (without imposing the conditions for the first-order phase transition) the model remains alive whether the singlet scalar stands for only a fraction or the whole dark matter relic abundance. The viable parameter space with couplings of order unity lies in the DM mass between the Higgs mass and $300$ GeV or above $1$ TeV where for the later the scalar field can constitute all the DM content. On the other hand, the real singlet scalar model is also capable of giving a first-order electroweak phase transition (EWPT) from the symmetric phase to the broken phase of the $SU(2)$ electroweak gauge symmetry group. Some papers that have studied also the phase transition in the single scalar model are \cite{McDonald:1993ey,Espinosa:2011ax,Ham:2004cf,Ahriche:2007jp,Huang:2015bta,Vaskonen:2016yiu, Chen:2017qcz, Beniwal:2017eik,Kurup:2017dzf,Kang:2017mkl, Chiang:2018gsn}. It has been pointed out that the dark matter constraints are strongly  in conflict with the first-order phase transition conditions (see e.g. \cite{Cline:2012hg}), while we have shown in \cite{Ghorbani:2018yfr} that in fact the observed dark matter relic density and the first-order electroweak phase transition are consistent, but the parameter space significantly gets reduced only after imposing the invisible Higgs decay constraint and gets completely excluded after considering the bounds from the direct detection experiments. 

It was shown in \cite{Ghorbani:2014gka} that by adding another real singlet scalar to the single scalar model, the problem of the restrictive direct detection constraints gets resolved non-trivially. The presence of a term $s s' H^\dagger H$ in the Lagrangian used in \cite{Ghorbani:2014gka}, with $s$ and $s'$ being the two real singlet scalars, is a key term in making the {\it two-scalar model} significantly different from simply summing up two single scalar models. The idea of two-scalar extension of the SM has been explored also in \cite{Casas:2017jjg} to address the DM and in \cite{Chao:2017vrq,Cheng:2018axr} for the Higgs inflation and the electroweak phase transition. Models with a complex scalar field or a composite Higgs have been studied in \cite{Branco:1998yk,Gonderinger:2012rd,Costa:2014qga,Jiang:2015cwa,Chala:2016ykx,Chiang:2017nmu,Grzadkowski:2018nbc,McDowall:2018tdg,Hektor:2019ote} and models with multi-scalar extension of the SM are found in \cite{Drozd:2011aa,Cheng:2018axr}.

In this paper we investigate in detail the question of the strongly first-order electroweak phase transition and the problem of dark matter simultaneously in the two-scalar model. Having two real scalars, in addition to the Higgs doublet scalar field $H$, the fields configuration space becomes three dimensional which in turn makes the structure of the phase transitions richer. Let us assume the vacuum expectation values (VEV) of the Higgs and the two extra scalars i.e. the VEVs of $(H, s, s')$, by $(v_\text{sym},w_1,w'_1)$ in the symmetric phase and $(v_\text{brk},w_2,w'_2)$ in the broken phase. In high temperatures that the electroweak symmetry group $SU(2)\times U(1)$ is not broken, the VEV of the Higgs is vanishing, therefore  throughout the paper we set $v_\text{sym}=0$. To stabilize the dark matter candidate, here chosen to be the scalar $s$, by a discrete symmetry group $\mathbb Z_2$, we need to set the VEV of the dark matter to zero after the phase transition, i.e. $w_2=0$. Therefore the general form of the transition from symmetric to broken phase is 
$(0,w_1,w'_1)\to (v_\text{brk},0,w'_2)$. We analyze different scenarios depending on possible values of $w_1, w'_1, v_\text{brk}$ and $w'_2$ to give a strongly first-order phase transition. We then combine the analytic conditions of the EWPT with the constraints from the direct and indirect dark matter searches. Despite the strong bounds from the first-order EWPT and the direct detection constraints which excludes completely the single scalar model, the two-scalar model evades remarkably all these constraints at the same time and predicts viable dark matter models. 

The paper is organized as follows. In Sec. \ref{first-order} we show analytically that there are different channels of the EWPT and obtain the necessary conditions for the EWPT to be of the first-order type. The section is divided into two subsection with two two-scalar models; one without the $s \mbox{-} s'$ cross-coupling terms and the other including these terms. Then in Sec. \ref{dmcons} we elaborate the DM relic density and direct detection constraints. In Sec. \ref{numres} we numerically search for the viable space of parameters combining the strongly first-order EWPT conditions, the observed dark matter relic density,  the direct detection constraints and the limit of the invisible Higgs decay width. We also compare the results with the single scalar model exposed to the first-order EWPT and the DM direct and indirect bounds. We conclude and summarize in Sec. \ref{conc}. In appendices \ref{minima} and \ref{crt} we bring the details of finding the minima of the scalars configuration $(H,s,s')$ and the deepest minimum condition respectively.

\section{First-Order Phase Transition}\label{first-order}
The strongly first-order phase transition in the early universe is one of the three Sakharov conditions \cite{Sakharov:1967dj} for the electroweak baryogenesis.  In high temperature of the early universe, the  electroweak symmetry group is unbroken and rest in its symmetric phase, $SU(2)\times U(1)$, with the Higgs VEV vanishing, but as the universe expanded, i.e. at lower temperatures, the vacuum acquires a non-vanishing VEV and the symmetry is broken into $U(1)$ electromagnetic guage group. In the SM framework, a strong first-order phase transition gives the Higgs mass an upper limit, $m_\text{H}<48$ GeV which is in conflict with the measured Higgs mass at the LHC being $125$ GeV. This motivates the extension of the SM  which among numerous possible extensions the addition of a real singlet scalar is the simplest. However as it has been shown in \cite{Ghorbani:2018yfr}, the viable space of parameters survived from the DM and the EWPT constraints, gets excluded mostly by taking into account the invisible Higgs decay constraint. Here we investigate the idea of extending the SM by two real scalars and examine the model against the simultaneous consideration of the DM and the EWPT along side the imposition of the direct and indirect probes.  In the two-scalar model we restrict ourselves to only terms with dimensionless couplings, therefore terms such as $s^3$ or $s s'^2$ are not present. Moreover, we analyze the model in two parts, once with the $s \mbox{-} s'$ cross-coupling terms for the scalars $s$ and $s'$, i.e. $s^2 s'^2$ and $s s'^3$ and $s^3 s'$, and once without these $s \mbox{-} s'$ cross-coupling terms. 

It can be seen from Eq. (\ref{pot}) that at very high temperature, $T\to \infty$, the only extremum of the thermal effective potential is the point $(v=0, w=0, w'=0)$ in VEV space. However, with the expansion of the universe as the temperature decreases, the non-zero local minima for the scalars come into existence. So in principle as the universe cools down  from very high to very low temperature, any of the scalar fields $h$, $s$ and $s'$ may undergo more than once a transition from a symmetric phase to a broken phase. In this paper by electroweak phase transition we mean the transition from a vanishing VEV into a non-zero VEV for the Higgs scalar field. We are not interested here in considering the scenarios of the symmetry breaking  in the dark sector. Therefore, in a symbolic transition from $(v_{\text{sym}},w_1,w'_1)\to (v_{\text{brk}},w_2,w'_2)$ in the VEV space, the parameters $w_1, w'_1$ are the VEVs of the scalars $s, s'$ at temperature $T_c + \delta T_1$ and $w_2, w'_2$ are the VEV's of the scalars at $T_c - \delta T_2$ for some arbitrary $\delta T_1$ and $\delta T_2$ and for $T_c$ being the critical temperature at which the phase transition in the Higgs sector triggers. The phase transition may continue until $T_c - \delta T_2$ approaches zero or it may end before the zero temperature. On the other hand, $\delta T_1$ can be arbitrarily small so that it is enough for the VEVs,  
$w_1$ and $w'_1$, to exist before or very close to the critical temperature. We will follow this strategy throughout the paper. 

\subsection{Model without $s \mbox{-} s'$ cross-coupling terms}\label{nnself}
The potential of the model possessing two extra real scalars beyond the SM along side the Higgs doublet, and without the $s \mbox{-} s'$ cross-coupling terms reads,

\begin{equation}\label{treepot}
\begin{split}
 V_{\text{tr}}(H,s,s')=&-\mu_{\text{h}}^{2}H^\dagger H+\lambda_{\text{h}}(H^\dagger H)^2\\
 &-\frac{1}{2}\mu_{\text{s}}^{2}s^{2}+\frac{1}{4}\lambda_{\text{s}}s^{4}
  -\frac{1}{2}\mu_{\text{s}'}^{2}s'^{2}+\frac{1}{4}\lambda_{\text{s}'}s'^{4}\\
 &+\lambda_{\text{hs}} s^{2}H^\dagger H+\lambda_{\text{hs}'} s'^{2}H^\dagger H
 +\lambda_{\text{hss}'} s s' H^\dagger H \,,
 \end{split}
 \end{equation}
where $H^\dagger=\frac{1}{\sqrt{2}}(0~v+h)$ denotes the Higgs doublet scalar field after the symmetry breaking, and $s,s'$ are the two real singlet scalar fields. The dominant one-loop thermal effective potential is given by \footnote{See \cite{Espinosa:1993bs} for one-loop thermal corrections in  potential with only one extra real singlet scalar.}, 
\begin{equation}\label{Vterloop}
 V_T^{\text{1-loop}}(h,s,s';T)\simeq \left( \frac{1}{2} c_\text{h} h^2 + \frac{1}{2} c_\text{s} s^2 + \frac{1}{2} c_{\text{s}'} s'^2 \right) T^2 \,,
\end{equation}
where 
\begin{subequations}\label{chcscsp}
\begin{align}
 &c_\text{h}=\frac{1}{48}\left( 9g^2 + 3g'^2 + 12y_t^2 + 12\lambda_\text{h} + 4\lambda_{\text{hs}} + 4\lambda_{\text{hs}'} \right)\,, \\
 &c_\text{s}= \frac{1}{12}\left(  3\lambda_\text{s} + \lambda_{\text{hs}}   \right)\,,\\
 &c_{\text{s}'}= \frac{1}{12}\left( 3\lambda_{\text{s}'} + \lambda_{\text{hs}'}  \right)\,.
 \end{align}
\end{subequations}

The thermal effective potential is obtained by summing up Eq. (\ref{treepot}) and Eq. (\ref{Vterloop}), 
\begin{equation}
 V_{\text{eff}}=V_{\text{tr}}(h,s,s')+V_T^{\text{1-loop}}(h,s,s';T)\,,
\end{equation}
that explicitly is given by Eq. (\ref{pot}) ignoring the cross-coupling terms. 
The potential is invariant under $\mathbb{Z}_2$ transformation if applied for both scalars at the same time, $s\to -s , s'\to - s'$. It means only through both scalars $s$ and $s'$ the $\mathbb{Z}_2$
symmetry is reserved, therefore the lighter scalar could be assumed as the dark matter candidate. 
In this paper, we take the scalar $s$ to be the dark matter particle.
The effect of the thermal correction is only in the mass term of the tree-level potential. So in the total effective potential instead of the coefficients $\mu_{\text{h}}^{2}, \mu_{\text{s}}^{2}$ and $\mu_{\text{s}'}^{2}$ we deal with $T$-dependent masses $\mu_{\text{h}}^{2}(T), \mu_{\text{s}}^{2}(T)$ and $\mu_{\text{s}'}^{2}(T)$ which are defined in (\ref{muhss}). 

The VEVs of the scalar fields $(h,s,s')$ can take different values before and after the EWPT. What is important to have in mind, is that after the EWPT, the VEV of the Higgs particle should be non-zero and the VEV of the lighter scalar field which is the DM candidate must be vanishing. Therefore, the most general structure of the VEVs would be $(v\neq 0, 0, w\neq 0)$. It is shown in appendix \ref{minima} that after the EWPT if we choose one of the scalar's VEV to be zero, the other scalar must take a vanishing VEV as well. So the only possibility for the VEVs after the EWPT is $(v\neq 0, 0, 0)$. 

\subsection*{Phase Transition Scenarios}

As seen in appendix \ref{minima}, in the symmetric phase where the Higgs vacuum expectation value is zero, there are four possibilities for  the two real scalars, $s$ and $s'$ to get zero or non-zero VEVs in order to solve the extremum conditions of the potential in Eq. (\ref{extermum}). As mentioned above, the set of VEVs for all the scalars after the EWPT has only one possibility: $(v^2=\frac{\mu^2_\text{h}(T)}{\lambda_\text{h}},0,0)$. Therefore in the model without the $s \mbox{-} s'$ cross-couplings, there can be four possible phase transitions i.e. from $(0,0,0)$, or $(0, w^2=\frac{\mu^2_{\text{s}}(T)}{\lambda_{\text{s}}},0)$, or $(0,0, w'^2=\frac{\mu^2_{\text{s}'}(T)}{\lambda_{\text{s}'}})$, or $(0, w^2=\frac{\mu^2_{\text{s}}(T)}{\lambda_s },w'^2=\frac{\mu^2_{\text{s}'}(T)}{\lambda_{\text{s}'} })$ to  $(v^2=\frac{\mu^2_\text{h}(T)}{\lambda_\text{h}},0,0)$ in which $\mu^2_{\text{h}}(T)$, $\mu^2_{\text{s}}(T)$ and $\mu^2_{\text{s}'}(T)$ are defined in Eq. (\ref{muhss}). Note these are only some selected solutions and in general there are more complicated expressions for $w^2$ and $w'^2$.  
We analyze all these four possible transitions one by one to figure out which can be of first order type. 

\subsubsection{Phase Transition \texorpdfstring{$(v=0, w=0,w'=0)\to (v\neq 0,w=0, w'=0)$}{Lg}}\label{a1}

In this scenario only the Higgs particle undergoes a non-zero VEV while the other two scalars keep the $\mathbb{Z}_2$ discrete symmetry in all low and high temperatures.  In order for $(0,0,0)$ to be a local minimum, Eq. (\ref{mincon}) must be satisfied. This would leave us with a set of constraints on $\mu^2(T)$'s, 
\begin{equation}\label{secder000}
  \mu^2_\text{h}(T)<0,~\mu^2_\text{s}(T)<0,~\mu^2_{\text{s}'}(T)<0\,,\\
\end{equation}
and a similar set of conditions must hold for $(v^2=\frac{\mu^2_\text{h}(T)}{\lambda_\text{h}},0,0)$, 
\begin{subequations}\label{secderv00}
\begin{align}
&\mu^2_\text{h}(T)>0 \label{secderv001}\,,\\
& - \mu^2_\text{s}(T)+\frac{\lambda_\text{hs}}{\lambda_\text{h}}\mu^2_\text{h}(T) >0\,, \\
&\left( - \mu^2_\text{s}(T)+\frac{\lambda_\text{hs}}{\lambda_\text{h}}\mu^2_\text{h}(T) \right) \left( -\mu^2_{\text{s}'}(T)+\frac{\lambda_{\text{hs}'}}{\lambda_\text{h}}\mu^2_\text{h}(T) \right)-\frac{1}{4} \frac{\lambda_{\text{hss}'}^2}{\lambda_\text{h}^2} \mu^4_\text{h}(T) >0\,.
\end{align}
\end{subequations}
The conditions on $\mu^2_\text{h}(T)$ in Eqs. (\ref{secder000}) and (\ref{secderv001}) are clearly inconsistent, which means that the two minima cannot coexist. Therefore, the first-order phase transition from $(0,0,0)$ to $(v,0,0)$ is not possible.

\subsubsection{Phase Transition \texorpdfstring{$(v=0,w\neq 0,w'=0)\to (v\neq 0, w=0, w'=0)$}{Lg}}\label{b1}

In appendix \ref{minima}, we see that $(v=0,w\neq 0,w'=0)$ with $w^2=\frac{\mu^2_{\text{s}}(T)}{\lambda_{\text{s}}}$ is an extremum of the potential. 
Here we examine the transition from $(v=0,w^2=\frac{\mu^2_{\text{s}}(T)}{\lambda_{\text{s}}},w'=0) $ to
$(v^2=\frac{\mu^2_{\text{h}}(T)}{\lambda_{\text{h}}}, w=0, w'=0)$. In other words, the mediator scalar, $s'$, gets always vanishing VEV before and after the phase transition. Then at high temperature, the Higgs has a zero VEV and the DM particle, $s$, has a non-zero VEV. This situation is closely related to the real single scalar dark matter model with the difference that here there is an additional real singlet scalar with a vanishing VEV. The minimum conditions for the point $(v\neq 0,w= 0,w'=0) $ is given in Eq. (\ref{secderv00}) and those for the VEV point $(v=0,w\neq 0,w'=0)$ can be extracted from Eq. (\ref{mincon}) in appendix \ref{minima},
\begin{subequations}\label{con0w0}
 \begin{align}
 & -\mu^2_\text{h}(T)+\frac{\lambda_\text{hs}}{\lambda_\text{s}}\mu^2_\text{s}(T) >0\,, \label{con0w01}\\
 & \mu^2_\text{s}(T)>0\,,\\
 &   \mu^2_{\text{s}'}(T)<0\,.
 \end{align}
\end{subequations}

The critical temperature below which the universe starts a transition from vanishing Higgs VEV
to non-zero VEV, is given by the following expressions, 
\begin{equation}\label{Tc0w0}
T^2_c=\frac{\mu^2_\text{s}-\sqrt{\frac{\lambda_\text{s}}{\lambda_\text{h}}}\mu^2_\text{h}}{c_s -\sqrt{\frac{\lambda_\text{s}}{\lambda_\text{h}}} c_\text{h}}\,,
\end{equation}
with $\lambda_\text{s}/\lambda_\text{h}>0$.

 For $T \leqslant T_c$ it is necessary that both $(0,w,0)$ and $(v,0,0)$ be local minima of the potential. Furthermore, the point $(v,0,0)$ in the VEV space must  be as well a global minimum for temperature below $T_c$. 
It can be shown that Eq. (\ref{con0w0}) holds for all values of the temperature in $0<T<T_c$, if it holds at $T=0$ and $T=T_c$, which leads to,
\begin{subequations}
 \begin{align}
& -\mu^2_\text{h}+\frac{\lambda_\text{hs}}{\lambda_\text{s}}\mu^2_\text{s}>0, &&-\mu^2_\text{h}+\frac{\lambda_\text{hs}}{\lambda_\text{s}}\mu^2_\text{s}+(c_\text{h}- \frac{\lambda_\text{hs}}{\lambda_\text{s}} c_s)T^2_c >0 \,, \\
&  \mu^2_\text{s}>0, &&\mu^2_\text{s}-c_\text{s} T^2_c>0 \,,\\
& \mu^2_{\text{s}'}<0,  && \mu^2_{\text{s}'}-c_{\text{s}'}T^2_c<0 \,.
 \end{align}
\end{subequations}
Then Eq. (\ref{secderv00}) holds for all $T \leqslant T_c$ if it holds only at $T=0$ and $T=T_c$, 
\begin{subequations}
 \begin{align}
 & \mu^2_\text{h}>0, \hspace{3.6cm} \mu^2_\text{h}-c_\text{h} T^2_c>0 \label{aaa}\,,\\
 & -\mu^2_\text{s}+\frac{\lambda_\text{hs}}{\lambda_\text{h}}\mu^2_\text{h}>0,\hspace{1.5cm} -\mu^2_\text{s}+\frac{\lambda_\text{hs}}{\lambda_\text{h}}\mu^2_\text{h}+\left( c_\text{s}-\frac{\lambda_\text{hs}}{\lambda_\text{h}}c_\text{h} \right) T^2_c >0 \,,\\
 & -\mu^2_{\text{s}'}+\frac{\lambda_{\text{hs}'}}{\lambda_\text{h}}\mu^2_\text{h}>0,\\
 & \left( - \mu^2_\text{s}(T_c)+\frac{\lambda_\text{hs}}{\lambda_\text{h}}\mu^2_\text{h}(T_c) \right) \left( -\mu^2_{\text{s}'}(T_c)+\frac{\lambda_{\text{hs}'}}{\lambda_\text{h}}\mu^2_\text{h}(T_c) \right)-\frac{1}{4} \frac{\lambda_{\text{hss}'}^2}{\lambda_\text{h}^2} \mu^4_\text{h}(T_c) >0\,.
 \end{align}
\end{subequations}

The last condition that must be considered in this scenario is that the minimum $(v,0,0)$ should be the global one for the temperatures below the critical temperature. That is, from Eq. (\ref{deltaV}) for all $T<T_c$, 
\begin{equation}
  \Delta V_\text{eff}\equiv V_\text{eff}(0,w,0;T)- V_\text{eff}(v,0,0;T)=-\frac{1}{4}\frac{\mu^4_\text{s}(T)}{\lambda_\text{s}}+\frac{1}{4} \frac{\mu^4_\text{h}(T)}{\lambda_\text{h}}>0\,.
\end{equation}
Equivalently, one can translate this constraint in $T^2$-derivative of $\Delta V_\text{eff}$ at $T=T_c$, 
\begin{equation}\label{globalmin0w0}
c_\text{s}-\sqrt{\frac{\lambda_\text{s}}{\lambda_\text{h}}}c_\text{h}<0\,,
\end{equation}
where use has been made of Eq. (\ref{con0w01}) and the following equality at $T=T_c$ from the definition of the critical temperature, 
\begin{equation}
 \frac{\mu^4_\text{s}(T_c)}{\lambda_\text{s}}=\frac{\mu^4_\text{h}(T_c)}{\lambda_\text{h}}\,.
\end{equation}

\subsubsection{Phase Transition \texorpdfstring{$(v=0,w= 0,w'\neq 0)\to (v\neq 0, w=0, w'=0)$}{Lg}}\label{c1}
This scenario is very similar to the last one, with the difference that here the DM candidate scalar, $s$, always takes zero VEV but the heavier scalar, $s'$,  goes from non-zero VEV before EWPT at high temperature to zero VEV at temperatures lower than the critical temperature.   
The local minimum conditions for the VEVs at low temperature after the EWPT, i.e. for $(v,0,0)$ are those given in Eq. (\ref{secderv00}). The conditions for  above the critical temperature are given by Eq. (\ref{con0w0}), but with an interchange in the scalar fields, i.e. $s\leftrightarrow s'$.
The critical temperature similarly is obtained, 
 \begin{equation}\label{Tc00w'}
T^2_c=\frac{\mu^2_{\text{s}'}-\sqrt{\frac{\lambda_{\text{s}'}}{\lambda_\text{h}}}\mu^2_\text{h}}{c_{s'} -\sqrt{\frac{\lambda_{\text{s}'}}{\lambda_\text{h}}} c_\text{h}}\,.
\end{equation}
For the VEV point $(v,0,0)$ to be the deepest minimum after the phase transition we have, 
\begin{equation}\label{globalmin00w'}
c_{\text{s}'}-\sqrt{\frac{\lambda_{\text{s}'}}{\lambda_\text{h}}}c_\text{h}<0\,.
\end{equation}

\subsubsection{Phase Transition \texorpdfstring{$(v=0,w\neq 0,w'\neq 0)\to (v\neq 0, w=0, w'=0)$}{Lg}}\label{d1}

In this scenario both scalars $s$ and $s'$ have non-zero VEVs above the critical temperature and both get zero VEV after the phase transition takes place. As is discussed in appendix \ref{crt}, the critical temperature can be obtained from Eq. (\ref{crtcon}), 
\begin{equation}
\begin{split}
 T^2_c=\frac{a\pm \sqrt{b \lambda_{\text{h}}\lambda_{\text{s}}\lambda_{\text{s}'}}}{c}\,,
 \end{split}
\end{equation}
where 
\begin{subequations}
 \begin{align}
  & a=  -c_\text{h}\lambda_\text{s}\lambda_{\text{s}'}\mu^2_\text{h}+c_\text{s}\lambda_\text{h}\lambda_{\text{s}'}\mu^2_\text{s}+c_{\text{s}'}\lambda_\text{h}\lambda_{\text{s}}\mu^2_{\text{s}'}\,, \\
  & b= \lambda_{\text{s}'}(c_\text{s}\mu^2_\text{h}-c_\text{h}\mu^2_\text{s})^2+c^2_{\text{s}'}(\lambda_{\text{s}}\mu^4_\text{h}-\lambda_\text{h}\mu^4_\text{s}) \nonumber \\
 & +2c_{\text{s}'}\mu^2_{\text{s}'}(-c_\text{h}\lambda_\text{s}\mu^2_\text{h}+c_\text{s}\lambda_\text{h}\mu^2_\text{s})+\mu^4_{\text{s}'}(-c^2_\text{s}\lambda_\text{h}+c^2_\text{h}\lambda_\text{s}) \,,\\
  & c= c^2_{\text{s}'}\lambda_\text{h}\lambda_\text{s}+c^2_{\text{s}}\lambda_\text{h}\lambda_{\text{s}'}-c^2_{\text{h}}\lambda_\text{s}\lambda_{\text{s}'}\,.
 \end{align}
\end{subequations}
The local minima conditions for the VEVs $(v=0, w\neq 0, w'\neq 0)$ before the EWPT are now more involved, 
\begin{subequations}\label{secder0ww'}
 \begin{align}
  & -\mu^2_\text{h}(T)+\frac{\lambda_{\text{hs}}}{\lambda_{\text{s}}}\mu^2_\text{s}(T)+\frac{\lambda_{\text{hs}'}}{\lambda_{\text{s}'}}\mu^2_{\text{s}'}(T)+\frac{\lambda_{\text{hss}'}}{\sqrt{\lambda_{\text{s}}\lambda_{\text{s}'}}}\sqrt{\mu^2_\text{s}(T)\mu^2_{\text{s}'}(T)}>0\,,\\
  &\mu^2_\text{s}(T)>0\,,\\
  &\mu^2_{\text{s}'}(T)>0\,,
 \end{align}
\end{subequations}
where must be satisfied at least for all $T\leqslant T_c$. These conditions at $T=0$ yields, 
\begin{subequations}\label{secder0ww'T0}
 \begin{align}
  & -\mu^2_\text{h}+\frac{\lambda_{\text{hs}}}{\lambda_{\text{s}}}\mu^2_\text{s}+\frac{\lambda_{\text{hs}'}}{\lambda_{\text{s}'}}\mu^2_{\text{s}'}+\frac{\lambda_{\text{hss}'}}{\sqrt{\lambda_{\text{s}}\lambda_{\text{s}'}}}\sqrt{\mu^2_\text{s}\mu^2_{\text{s}'}}>0 \,,\\
  &\mu^2_\text{s}>0\,,\\
  &\mu^2_{\text{s}'}>0\,,
 \end{align}
\end{subequations}
and at $T=T_c$, 
\begin{subequations}\label{secder0ww'Tc}
 \begin{align}
  &-\mu^2_\text{h}(T_c)+\frac{\lambda_{\text{hs}}}{\lambda_{\text{s}}}\mu^2_\text{s}(T_c)+\frac{\lambda_{\text{hs}'}}{\lambda_{\text{s}'}}\mu^2_{\text{s}'}(T_c)+\frac{\lambda_{\text{hss}'}}{\sqrt{\lambda_{\text{s}}\lambda_{\text{s}'}}}\sqrt{\mu^2_\text{s}(T_c)\mu^2_{\text{s}'}(T_c)}>0\,,\\
  &\mu^2_\text{s}-c_s T_c^2>0\,,\\
  &\mu^2_{\text{s}'}-c_{s'}T^2_c>0\,.
 \end{align}
\end{subequations}
In order to have a first order transition from symmetric phase to broken symmetry phase of the Higgs vacuum, the VEV set $(v,0,0)$ must be a global minimum. This condition is obtained via Eqs. (\ref{deltaV}) and is given by, 
\begin{equation}
 \frac{c_\text{s}}{\lambda_\text{s}}(\mu^2_\text{s}-c_\text{s} T^2_c)+\frac{c_{\text{s}'}}{\lambda_{\text{s}'}}(\mu^2_{\text{s}'}-c_{\text{s}'} T^2_c)-\frac{c_\text{h}}{\lambda_\text{h}}(\mu^2_\text{h}-c_\text{h} T^2_c)<0\,.
\end{equation}

\subsection{Model including $s \mbox{-} s'$ cross-coupling terms}\label{selfint}

In the previous subsection, we ignored the $s \mbox{-} s'$ cross-coupling terms, i.e. the interaction terms consisting only the singlet scalars, $s$ and $s'$. If we include also these terms, the total tree-potential would be the sum of the potential in Eq. (\ref{treepot}) and the $s \mbox{-} s'$ cross-coupling terms,  

\begin{equation}\label{Vself}
V = V_{\text{tr}}  +\frac{1}{2}\lambda_{\text{ss}'} s^2 s'^2 + \frac{1}{3}\lambda'_{\text{ss}'} s s'^3 + \frac{1}{3}\lambda''_{\text{ss}'} s^3 s'\,.
\end{equation}
Note that we have considered only the cross-coupling terms with dimensionless couplings. The one-loop thermal potential in this scenario has the same form as in Eq. (\ref{Vterloop}), however the coefficients $c_\text{h}$, $c_\text{s}$ and $c_{\text{s}'}$ are now different from Eqs. (\ref{chcscsp}) as now there are more one-loop Feynman diagrams for thermal mass corrections,
\begin{subequations}
\begin{align}
 &c_\text{h}=\frac{1}{48}\left( 9g^2 + 3g'^2 + 12y_t^2 + 12\lambda_\text{h} + 4\lambda_{\text{hs}} + 4\lambda_{\text{hs}'} \right)\,, \\
 &c_\text{s}= \frac{1}{12}\left(3\lambda_\text{s} + \lambda_{\text{hs}} +\lambda_{\text{ss}'}  \right)\label{csself}\,,\\
 &c_{\text{s}'}= \frac{1}{12}\left( 3\lambda_{\text{s}'} +\lambda_{\text{hs}'}+\lambda_{\text{ss}'}   \right)\label{cspself}\,.
 \end{align}
\end{subequations}
As seen in Eqs. (\ref{csself}) and (\ref{cspself}), the coupling $\lambda_{\text{ss}'}$ appears in the thermal corrections. The reason is that the one-loop thermal mass correction for the scalar $s$ ($s'$), in addition to the Higgs field, includes as well the scalar $s'$ ($s$) in the loop. However, the  couplings $\lambda'_{\text{ss}'}$ and $\lambda''_{\text{ss}'}$, although playing a role in first-order phase transition conditions, but they do not enter directly in the mass thermal corrections.

\subsection*{Phase Transition Scenarios}
Finding a complete set of the extrema $(v,w,w')$ (with $v$, $w$ and $w'$ being the VEV's of $h$, $s$ and $s'$ respectively) from Eqs. (\ref{eqset}), for the general case of totally non-vanishing $\lambda_{\text{ss}'}$, $\lambda'_{\text{ss}'}$ and $\lambda''_{\text{ss}'}$, is possible but the solutions are very lengthy. Therefore, in the following subsections we consider only the solutions which with no loss of generality are simpler and can also be compared with the phase transitions in model without $s \mbox{-} s'$ cross-coupling  terms.

\subsubsection{Phase Transition \texorpdfstring{$(v=0,w= 0,w'\neq 0)\to (v\neq 0, w=0, w'=0)$}{Lg}}\label{a2}
The phase transition here is from $(v=0,w= 0,w'^2=\frac{\mu^2_{\text{s}'}(T)}{\lambda_{\text{s}'}})$ to $(v^2=\frac{\mu^2_{\text{h}}(T)}{\lambda_{\text{h}}},w= 0,w'=0)$. 
This extremum solution of the potential is possible for at least two different choices of the $s \mbox{-} s'$ cross-couplings, i.e. for 
$\left( \lambda_{\text{ss}'}\neq 0 \,,\lambda'_{\text{ss}'}=0 \,, \lambda''_{\text{ss}'}= 0 \right) $ and for $\left(\lambda_{\text{ss}'}\neq 0 \,,\lambda'_{\text{ss}'} \neq 0 \,, \lambda''_{\text{ss}'}= 0 \right)$. Here we derive the first-order phase transition conditions for the first set of the $s \mbox{-} s'$ cross-couplings above which turns out to be the same as the other set of coupling. The local minimum conditions using Eqs. (\ref{secder}) and (\ref{mincon}) are, 

\begin{subequations}\label{00w'self}
 \begin{align}
  -\mu^2_\text{h}(T)+\frac{\lambda_{\text{hs}'}}{\lambda_{\text{s}'}}\mu^2_{\text{s}'}(T) >0\,,\\
  -\mu^2_\text{s}(T)+\frac{\lambda_{\text{ss}'}}{\lambda_{\text{s}'}}\mu^2_{\text{s}'}(T) >0\,,\\
  \mu^2_{\text{s}'}(T)>0\,,
 \end{align}
\end{subequations}
where similar to the lines in Sec. \ref{nnself}, it is enough that Eqs. (\ref{00w'self}) satisfy for $T=0$ and $T=T_c$, 
\begin{subequations}
 \begin{align}
  & -\mu^2_\text{h}+\frac{\lambda_{\text{hs}'}}{\lambda_{\text{s}'}}\mu^2_{\text{s}'}>0\,,&& -\mu^2_\text{h}+\frac{\lambda_{\text{hs}'}}{\lambda_{\text{s}'}}\mu^2_{\text{s}'}+\left(c_\text{h}-\frac{\lambda_{\text{hs}'}}{\lambda_{\text{s}'}}c_{\text{s}'} \right)T^2_c>0 \,,\\
  & -\mu^2_\text{s}+\frac{\lambda_{\text{ss}'}}{\lambda_{\text{s}'}}\mu^2_{\text{s}'} >0\,,&&-\mu^2_\text{s}+\frac{\lambda_{\text{ss}'}}{\lambda_{\text{s}'}}\mu^2_{\text{s}'} >0+ \left(c_\text{s}-\frac{\lambda_{\text{ss}'}}{\lambda_{\text{s}'}}c_{\text{s}'} \right) T^2_c >0 \,,\\
  & \mu^2_{\text{s}'}>0\,,&& \mu^2_{\text{s}'} - c_{\text{s}'}T^2_c  >0 \,.
 \end{align}
\end{subequations}
Similarly the local minimum conditions for the VEV set $(v,0,0)$ with $v^2=\frac{\mu^2_\text{h}}{\lambda_\text{h}}$, must be driven from Eqs. (\ref{secder}) and (\ref{mincon}). It turns out that these conditions for the model with  the $s \mbox{-} s'$ cross-coupling terms is the same as those for the model without the $s \mbox{-} s'$ cross-coupling terms in subsection \ref{nnself}, i.e, in Eqs. (\ref{secderv00}). 

The critical temperature is obtained from the degeneracy condition in Eq. (\ref{crtcon}) and is given by, 
\begin{equation}
T^2_c=\frac{\mu^2_{\text{s}'}-\sqrt{\frac{\lambda_{\text{s}'}}{\lambda_\text{h}}}\mu^2_\text{h}}{c_{s'} -\sqrt{\frac{\lambda_{\text{s}'}}{\lambda_\text{h}}} c_\text{h}}\,. 
\end{equation}
After the phase transition, the minimum in the broken phase needs to be a global minimum which is translated into, 
\begin{equation}
c_{\text{s}'}-\sqrt{\frac{\lambda_{\text{s}'}}{\lambda_\text{h}}}c_\text{h}<0\,.
\end{equation}
\subsubsection{Phase Transition \texorpdfstring{$(v=0,w\neq 0,w'= 0)\to (v\neq 0, w=0, w'=0)$}{Lg}}\label{b2}
In this scenario the phase transition is from $(v=0,w=\frac{\mu^2_{\text{s}}(T)}{\lambda_{\text{s}}},w'= 0)\to (v=\frac{\mu^2_{\text{h}}(T)}{\lambda_{\text{h}}}, w=0, w'=0)$. It means that the DM candidate takes non-zero VEV before the phase transition and its VEV flips to zero after the phase transition to retain the $\mathbb{Z}_2$ symmetry. Again there are two sets of the $s \mbox{-} s'$ cross-couplings for which the VEV set before the phase transition is an extremum solution to the potential in Eq. (\ref{Vself}): $\left( \lambda_{\text{ss}'}\neq 0\,, \lambda'_{\text{ss}'}= 0 \,,\lambda''_{\text{ss}'} =0 \right) $ and for $\left(\lambda_{\text{ss}'}\neq 0\,, \lambda'_{\text{ss}'}= 0 \,,\lambda''_{\text{ss}'} \neq 0 \right)$. For both sets of the couplings, the local minimum conditions for the point $(v=0,w=\frac{\mu^2_{\text{s}}(T)}{\lambda_{\text{s}}},w'= 0)$ is given by, 
\begin{subequations}\label{0w0self}
 \begin{align}
 & -\mu^2_\text{h}(T)+\frac{\lambda_{\text{hs}}}{\lambda_{\text{s}}}\mu^2_{\text{s}}(T) >0\,,\\
 & \mu^2_{\text{s}}(T)>0\,,\\
 & -\mu^2_{\text{s}'}(T)+\frac{\lambda_{\text{ss}'}}{\lambda_{\text{s}}}\mu^2_{\text{s}}(T) >0\,,
 \end{align}
\end{subequations}
which is held for all $T\leqslant T_c$ if, 
\begin{subequations}
 \begin{align}
 & -\mu^2_\text{h}+\frac{\lambda_{\text{hs}}}{\lambda_{\text{s}}}\mu^2_{\text{s}}>0\,,&& -\mu^2_\text{h}+\frac{\lambda_{\text{hs}}}{\lambda_{\text{s}}}\mu^2_{\text{s}}+\left(c_\text{h}-\frac{\lambda_{\text{hs}}}{\lambda_{\text{s}}}c_{\text{s}} \right)T^2_c>0 \,,\\
& \mu^2_{\text{s}}>0\,,&& \mu^2_{\text{s}} - c_{\text{s}}T^2_c  >0 \,,\\
  & -\mu^2_{\text{s}'}+\frac{\lambda_{\text{ss}'}}{\lambda_{\text{s}}}\mu^2_{\text{s}} >0\,,&& -\mu^2_{\text{s}'}+\frac{\lambda_{\text{ss}'}}{\lambda_{\text{s}}}\mu^2_{\text{s}} >0+ \left(c_{\text{s}'}-\frac{\lambda_{\text{ss}'}}{\lambda_{\text{s}}}c_{\text{s}} \right) T^2_c >0 \,,
 \end{align}
\end{subequations}
where $T_c$, the critical temperature, for this scenario is given by, 
\begin{equation}
T^2_c=\frac{\mu^2_{\text{s}}-\sqrt{\frac{\lambda_{\text{s}}}{\lambda_\text{h}}}\mu^2_\text{h}}{c_{s} -\sqrt{\frac{\lambda_{\text{s}}}{\lambda_\text{h}}} c_\text{h}}\,. 
\end{equation}
In order for the minimum $(v,0,0)$ to be deeper than  $(0,w,0)$, the following condition must be held, 
\begin{equation}
c_{\text{s}}-\sqrt{\frac{\lambda_{\text{s}}}{\lambda_\text{h}}}c_\text{h}<0\,,
\end{equation}
where the degeneracy condition on the potential at the critical temperature has been used. Again, the local minimum conditions for $(v,0,0)$ is given by Eqs. (\ref{secderv00}).

\subsubsection{Phase Transition \texorpdfstring{$(v=0,w\neq 0,w'\neq 0)\to (v\neq 0, w=0, w'=0)$}{Lg}}\label{c2}
This type of the phase transition from $(v=0,w\neq 0,w'\neq 0)$ to $(v\neq 0, w=0, w'=0)$, as seen in the appendix \ref{selfcase}, is possible for a choice of the $s \mbox{-} s'$ cross-couplings being $\left( \lambda_{\text{ss}'}\neq 0\,, \lambda'_{\text{ss}'}= 0 \,,\lambda''_{\text{ss}'} =0 \right)$ with $w$ and $w'$ given by, 
\begin{equation}\label{0ww'self}
w^2= \frac{\lambda_{\text{s}'} \mu^2_{\text{s}}(T)-\lambda_{\text{ss}'} \mu^2_{\text{s}'}(T)}{\lambda_{\text{s}}\lambda_{\text{s}'}-\lambda_{\text{ss}'}^2},~~~ w'^2= \frac{\lambda_{\text{s}} \mu^2_{\text{s}'}(T)-\lambda_{\text{ss}'} \mu^2_{\text{s}}(T)}{\lambda_{\text{s}}\lambda_{\text{s}'}-\lambda_{\text{ss}'}^2}\,.
\end{equation}
The local minimum conditions for the point $(0,w,w')$ in the VEV space, is obtained from Eqs. (\ref{secder}) and \ref{mincon}, 
\begin{subequations}
 \begin{align}
 \begin{split}
  & -\mu^2_{\text{h}}(T)+  \left(\lambda_{\text{s}} \lambda_{\text{s}'}-\lambda_{\text{ss}'}^2\right)^{-1} \\
  & \times \Big[ \lambda_{\text{hs}'}  \left(\lambda_{\text{s}} \mu^2_{\text{s}'}(T)-\lambda_{\text{ss}'} \mu^2_{\text{s}}(T) \right)+\lambda_{\text{hs}} \left(\lambda_{\text{s}'} \mu^2_{\text{s}}(T)-\lambda_{\text{ss}'} \mu^2_{\text{s}'}(T) \right) \\
& +  \lambda_{\text{hss}'} \sqrt{\left(\lambda_{\text{s}} \mu^2_{\text{s}'}(T)-\lambda_{\text{ss}'} \mu^2_{\text{s}}(T) \right) \left(\lambda_{\text{s}'} \mu^2_{\text{s}}(T)-\lambda_{\text{ss}'} \mu^2_{\text{s}'}(T) \right) } \Big] >0 \,,
  \end{split}\\
  \begin{split}
 & -\mu^2_{\text{s}}(T)+   \left(\lambda_{\text{s}} \lambda_{\text{s}'}-\lambda_{\text{ss}'}^2\right)^{-1} \\
 & \times \Big[\lambda_{\text{ss}'} \left(\lambda_{\text{s}} \mu^2_{\text{s}'}(T)-\lambda_{\text{ss}'} \mu^2_{\text{s}}(T) \right)+3 \lambda_{\text{s}}  \left(\lambda_{\text{s}'} \mu^2_{\text{s}}(T)-\lambda_{\text{ss}'} \mu^2_{\text{s}'}(T) \right) \Big] >0 \,,
  \end{split}\\
  \begin{split}
& \frac{ \left(\lambda_{\text{s}} \mu^2_{\text{s}'}(T)-\lambda_{\text{ss}'} \mu^2_{\text{s}}(T) \right)\left(\lambda_{\text{s}'} \mu^2_{\text{s}}(T)-\lambda_{\text{ss}'} \mu^2_{\text{s}'}(T) \right)}{\lambda_{\text{s}} \lambda_{\text{s}'}-\lambda_{\text{ss}'}^2} >0 \,.
\end{split}
 \end{align}
\end{subequations}
It is enough that Eqs. (\ref{0ww'self}) satisfy at $T=0$ and $T=T_c$ in order to hold for all $T\leqslant T_c$. The local minimum conditions for the point $(v,0,0)$ with $v^2=\frac{\mu^2_\text{h}(T)}{\lambda_\text{h}}$ is given by Eq. (\ref{secderv00}). 
The critical temperature reads,
\begin{equation}
 T^2_c= \frac{\lambda_{\text{h}} \left( c_{\text{s}'} b_1 + c_{\text{s}} b_2 \right)
 - c_{\text{h}} \mu^2_{\text{h}} a
 + \sqrt{  - \lambda_{\text{h}} a \left( \lambda_{\text{s}'} c^2_3 + c^2_{\text{s}'} c_1 +2 c_{\text{h}}\lambda_{\text{ss}'} \mu^2_{\text{s}'} c_3 + \mu^4_{\text{s}'} c_2 + 2 c_{\text{s}'} d_1 \right)} 
 } 
 {d_2-c^2_{\text{h}} a}
\end{equation}
where
\begin{equation}\label{abcd}
 \begin{split}
 &a=\lambda_{\text{s}} \lambda_{\text{s}'}-\lambda_{\text{ss}'}^2 \,, \\
 & b_1= \lambda_{\text{s}} \mu^2_{\text{s}'} - \lambda_{\text{ss}'} \mu^2_{\text{s}} \,, \\
 &  b_2= \lambda_{\text{s}'} \mu^2_{\text{s}} - \lambda_{\text{ss}'} \mu^2_{\text{s}'} \,, \\
 &  c_1= \lambda_{\text{s}} \mu^4_{\text{h}} -\lambda_{\text{h}} \mu^4_{\text{s}} \,, \\
  &   c_2= \lambda_{\text{s}} c^2_{\text{h}} -\lambda_{\text{h}} c^2_{\text{s}} \,, \\
 &     c_3= c_{\text{s}} \mu^2_{\text{h}} - c_{\text{h}} \mu^2_{\text{s}} \,, \\
 &  d_1=  \lambda_{\text{ss}'} \left( -c_{\text{s}} \mu^4_{\text{h}} + c_{\text{h}}  \mu^2_{\text{h}}\mu^2_{\text{s}} \right)
  + \mu^2_{\text{s}'} \left( - c_{\text{h}} \lambda_{\text{s}}  \mu^2_{\text{h}}  + c_{\text{s}} \lambda_{\text{h}}  \mu^2_{\text{s}} \right) \,, \\
 & d_2= \lambda_{\text{h}} \left( c^2_{\text{s}'} \lambda_{\text{s}} + c^2_{\text{s}} \lambda_{\text{s}'} - 2  c_{\text{s}} c_{\text{s}'} \lambda_{\text{ss}'}  \right) \,.
 \end{split}
\end{equation}

Now the minimum $(v,0,0)$ after the electroweak symmetry breaking must be deepest  minimum and therefore Eq. (\ref{deltaV}) must be satisfied, 
\begin{equation}
 \Delta V_\text{eff}(T)\equiv V_\text{eff}(0,w,w';T)- V_\text{eff}(v_\text{brk},0,0;T)>0
\end{equation}
where $v^2=\frac{\mu^2_\text{h}(T)}{\lambda_\text{h}}$ and $w, w'$ are given by Eq. (\ref{0ww'self}). The above inequality leads to the following condition, 
\begin{equation}
 -\frac{c_{\text{h}}}{\lambda_{\text{h}}} \mu^2_{\text{h}}(T_c)+\frac{\lambda_{\text{s}}}{a}\mu^2_{\text{s}'}(T_c)+\frac{\lambda_{\text{s}'}}{a}\mu^2_{\text{s}}(T_c)-\frac{\lambda_{\text{ss}'}}{a}\left( c_{\text{s}} \mu^2_{\text{s}'}+c_{\text{s}'} \mu^2_{\text{s}}-2 c_{\text{s}}c_{\text{s}'} T^2_c \right)<0\,,
\end{equation}
where $a$ is the parameter defined in Eq. (\ref{abcd}).

One of the Sakharov conditions for the baryogenesis is the washout criterion which guarantees an appropriate sphaleron rate to have a strongly first-order phase transition. This condition is translated into an inequality as $v_c/T_c>1$. In all the numerical computation and for each phase transition scenario, we impose also the washout criterion.

\section{Dark Matter Constraints}\label{dmcons}
In the last section we elaborately studied the simplest possible phase transitions that may occur for going from the symmetric phase to the broken phase of the Higgs vacuum. In this section we discuss the dark matter constraints for the two-scalar model and in the next section we combine these constraints with those of the strongly first-order phase transition in the last section and represent the final results. 

From the last section, having observed that the only possibility for the VEVs of the scalars after the EWPT is the VEV structure $(v,0,0)$, let us study the mixing among the scalars after the electroweak symmetry breaking. The mass matrix after the Higgs particle gets its non-zero VEV is not diagonal, although both scalars $s$ and $s'$ are at their zero VEVs.  Diagonalizing the mass matrix can be done by a rotation in the $(s,s')$ field configuration, 
\begin{equation}
 \left( \begin{matrix}
   s \\
  s'
 \end{matrix}
 \right)
\rightarrow
\left( \begin{matrix}
   \phi \\
  \phi'
 \end{matrix}
 \right)\equiv 
 \left( \begin{matrix}
   s \cos\theta + s' \sin\theta \\
  -s \sin\theta + s' \cos\theta 
 \end{matrix}
 \right)
\end{equation}
where $\phi$ and $\phi'$ are new scalars with diagonalized mass eigenvalues, and $\theta$ is the mixing angle and is given by, 

\begin{equation}\label{tan}
 \tan{2\theta }=\frac{2 v^2 \lambda_{\text{hss}'}}{m_{\text{s}'}^2-m_{\text{s}}^2}
\end{equation}
in which,
\begin{subequations}
\begin{align}
 &m_\text{s}^2=v^2 \lambda_{\text{hs}}- \mu_\text{s}^2\\
 &m_{\text{s}'}^2=v^2 \lambda_{\text{hs}'}-\mu_{\text{s}'}^2
 \end{align}
\end{subequations}
are the masses of the scalars $s$ and $s'$ which are related to the VEV of the Higgs scalar, $v$.
The extremum condition of the potential in eq. (\ref{treepot}) at $(v,0,0)$ gives $\mu_\text{h}^2=v^2 \lambda_\text{h}$.  \\
The entries of the diagonalized mass matrix after the rotation from the field configuration $(s,s')$ into $(\phi,\phi')$ read, 
\begin{subequations}
\begin{align}
&  m_\text{h}^2=2 v^2 \lambda_\text{h} \\
 &m_\phi^2=- v^2 \lambda_{\text{hs}} 
 + m_\text{s}^2 \cos^2\theta + m_{\text{s}'}^2 \sin^2\theta\\
 & m_{\phi'}^2= -v^2 \lambda_{\text{hs}'} 
 + m_\text{s}^2 \sin^2\theta + m_{\text{s}'}^2 \cos^2\theta
 \end{align}
\end{subequations}
Therefore, in two-scalar model we have two singlet scalar WIMPs, $\phi$ and $\phi'$ where we assume the DM candidate is the $\phi$ field being the stable WIMP, and the heavier WIMP, $\phi'$, is unstable and can decay to the SM particles plus the light WIMP through an intermediate Higgs, i.e. $\phi'\to \phi+ \text{SM}$. Let us define the mass difference of the two scalars as $\delta = m_{\phi'} - m_{\phi}$. For the mass splitting of ${\cal O} (\text{GeV})$ and beyond, the life time of the heavy WIMP will be much smaller than the age of the universe and therefore cannot have effective contribution to the present DM relic density \cite{Ghorbani:2014gka}. 
One of the important constraints on the DM models comes from 
the observed DM relic density given by WMAP and Planck experiments. 
In this work we assume the present DM relic density is produced thermally
via the so-called freeze-out mechanism taken place at some specific 
freeze-out temperature, $T_f$,  in the early universe \cite{Lee:1977ua}. 
Let us discuss briefly how this mechanism works. 

At high temperatures we believe that the WIMPs and 
the SM particles are in thermal equilibrium in an expanding universe. 
It means that the WIMPs annihilation into the SM particles and, the WIMPs productions
take place at the same rate. 
As the universe expands, there is an epoch after which the universe 
expansion rate surpasses the WIMPs annihilation rate such that it becomes
infrequent for the WIMPs to meet each other for the annihilation to happen. 
At this point in time, the temperature is low enough so that the  
SM particles possess insufficient kinetic energy to produce WIMPs. This is the epoch in which DM particles decouple from the SM particles, thenceforth the DM density remains constant asymptotically in the 
comoving volume. 

\begin{figure}
\begin{center}
\includegraphics[width=.8\textwidth,angle =0]{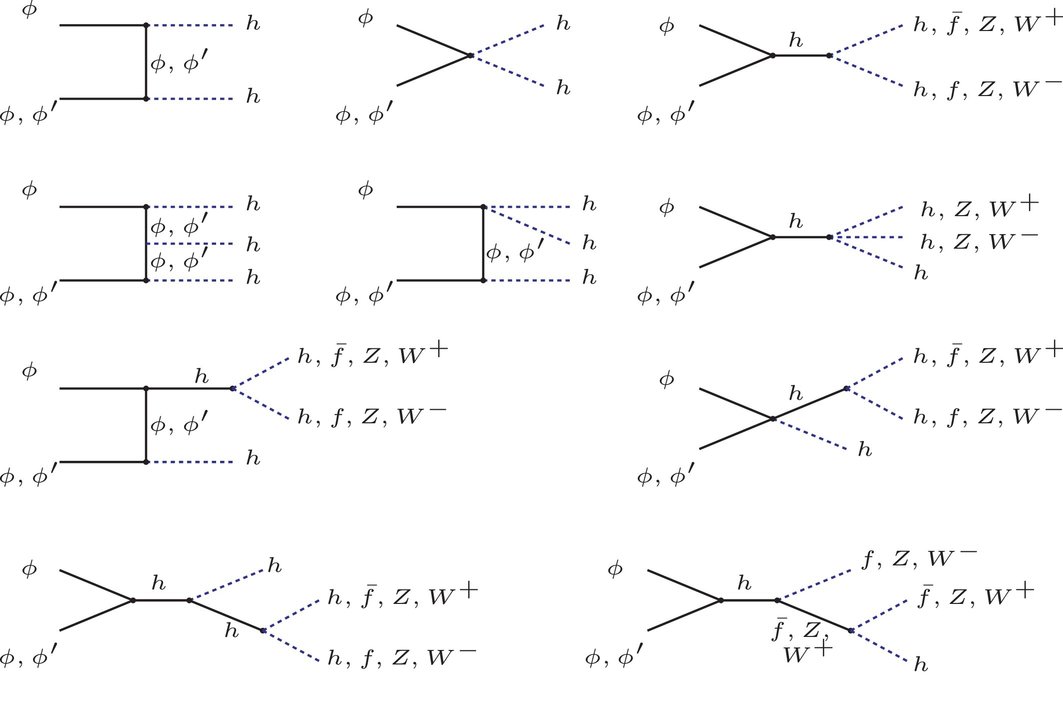}
\end{center}
\caption{Annihilation and coannihilation Feynman diagrams are shown 
up to three particles in the final state.} 
\label{anni-diagrams}
\end{figure}

The freeze-out temperature and hence the DM relic density 
depend strongly on various types of the WIMP interactions with the SM particles. 
The dominant contributions are due to DM annihilation cross section and the less important contributions 
come from coannihilation processes. 
In the later case, the DM candidate along with 
the heavier WIMP annihilate into the SM particles.
The relevant (co)annihilation Feynman diagrams for the two-scalar model up to three particles in the final state are shown in Fig.~\ref{anni-diagrams}.
When two particles in the final state, the DM (co)annihilations can proceed in three different ways. 
The (co)annihilations to two SM Higgs can be through $t$- and $u$-channels with an intermediate DM or an intermediate heavier WIMP. 
The second type of the process is the (co)annihilation to all the SM particles except the neutrinos via the SM Higgs in the $s$-channel.
The last possible way for DM (co)annihilation is a contact interaction with two Higgs in the final state. 

The change in the number densities of the two WIMPs in terms of the temperature 
are controlled by two coupled Boltzmann equations. 
Instead of solving the two coupled equations which is not a simple task, one can solve a single Boltzmann equation with an effective 
DM cross section incorporating both annihilation and coannihilation cross sections \cite{Griest:1990kh,Edsjo:1997bg}.
If we take the total number density as $n=n_{\phi} + n_{\phi'}$, 
the effective Boltzmann equation reads,

\begin{equation}
 \frac{dn}{dt}=-3Hn-\braket{\sigma_{\text{eff}}\,v}\left(n^2 - n^2_{\text{eq}} \right)\, ,
\end{equation}
where the effective cross section is defined as 
\begin{equation}
\sigma_{\text{eff}}=\frac{1}{g_{\text{eff}}}\left(\sigma_{\phi\phi}+\sigma_{\phi'\phi'}\left(1+\frac{\delta}{m_{\phi}}\right)^{3}e^{-2\delta/T}
+2\sigma_{\phi\phi'}\left(1+\frac{\delta}{m_{\phi}}\right)^{3/2}e^{-\delta/T} \right)\,.
\end{equation}
Here, $\sigma_{\phi\phi}$, $\sigma_{\phi'\phi'}$ and $\sigma_{\phi\phi'}$ 
denote respectively, the DM annihilation to the SM particles, the heavier 
WIMP annihilation to the SM particles and the coannihilation to the SM particles. 
The effective number of degrees of freedom is $g_{\text{eff}} = 1+(1+\delta/m_{\phi})^{3/2} e^{-\delta/T}$, and the Hubble constant in the Boltzmann equation is denoted by $H$. The thermal averaging of the effective cross section multiplied by the relative DM velocity at temperature $T$ is defined as 
$\braket{\sigma_{\text{eff}}\,v}$. 

\begin{figure}
\begin{center}
\includegraphics[width=.18\textwidth,angle =0]{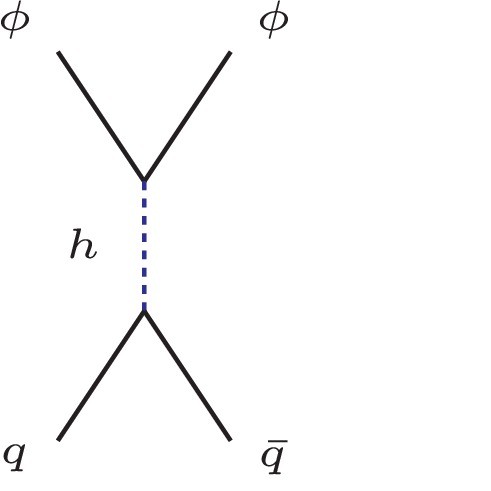}
\end{center}
\caption{The DM-quark direct detection scattering cross section is shown
         at the leading order in perturbation theory.} 
\label{DD-diagram}
\end{figure}

The second important constraint that one should impose on the parameter space is the stringent
exclusion limits from the DM-nucleon elastic scattering cross sections. 
These limits are provided by dark matter direction detection experiments 
among them we exploit here the latest updates of LUX \cite{Akerib:2016vxi} and XENON1T \cite{Aprile:2017iyp}. 
The underlying interaction which leads to DM-nucleon elastic scattering
is given by an effective Lagrangian describing the DM-quark interaction, 
\begin{equation}
{\cal L}_{\text{eff}} = c_{\text{q}} \phi \phi~ \bar q q \,, 
\end{equation}
where the effective coupling $c_{\text{q}}$ is obtained in terms of the relevant 
couplings in the Lagrangian, the mixing angle, the quark mass, and the Higgs mass as follows,
\begin{equation}
c_{\text{q}} = \frac{m_{\text{q}}}{m_{\text{h}}^2} 
(\lambda_{\text{hs}}\cos^2 \theta+\lambda_{\text{hs}'}\sin^2 \theta - \frac{1}{2} \lambda_{\text{hss}'} \sin 2\theta ).
\end{equation}
This type of interaction, results in a spin-independent (SI) DM-nucleon elastic 
scattering cross section. The DM-quark interaction in terms of Feynman diagram
is shown in Fig.~\ref{DD-diagram}.

There is a standard method by which one can promote the quark-level effective Lagrangian to hadron-level interaction at zero-momentum transfer \cite{Ellis:2008hf,Crivellin:2013ipa}.
This can be achieved if we replace the quark current by a nucleon current
up to a low energy effective factor $c_{\text{N}}$ as,
\begin{equation}
 c_{\text{N}} = m_{\text{N}} \left( \sum_{q = u,d,s} f^{\text{N}}_{Tq} \frac{c_{\text{q}}}{m_{\text{q}}} 
+ \frac{2}{27} f^{\text{N}}_{Tg} \sum_{q = c,b,t}   \frac{c_{\text{q}}}{m_{\text{q}}} \right) \,.
\end{equation}
For the DM-proton scattering cross section we use these scalar couplings, 
$f^{p}_{u} = 0.0153$, $f^{p}_{d} = 0.0191$, $f^{p}_{s} = 0.0447$ and $f^{p}_{g} = 1 - f^{p}_{u}-f^{p}_{d}-f^{p}_{s}$ \cite{Belanger:2013oya}.
The final formula for the DM-proton SI elastic scattering cross section is, 
\begin{equation}
 \sigma^{\text{p}}_{\text{SI}} = 
\frac{c_{p}^2 \mu_{p}^2}{\pi m_{\phi}^2}\,,
\end{equation}
where $\mu_{p}$ is the reduced mass of the DM and the proton.

\begin{figure}
\begin{subfigure}
 \centering
\includegraphics[width=.38\textwidth,angle =-90]{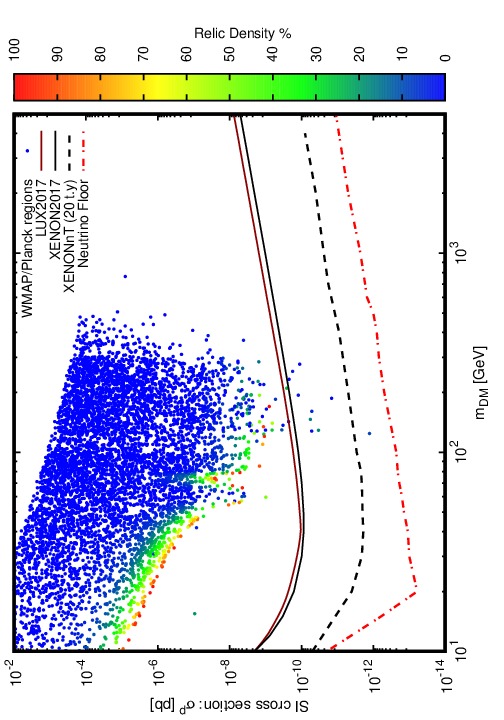}
\end{subfigure}
\begin{subfigure}
\centering
\includegraphics[width=.38\textwidth,angle =-90]{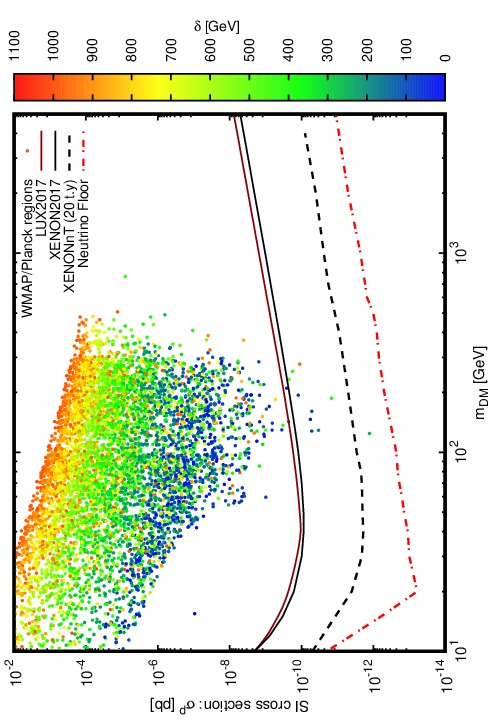}
\end{subfigure}
\newline
\begin{subfigure}
\centering
\includegraphics[width=.38\textwidth,angle =-90]{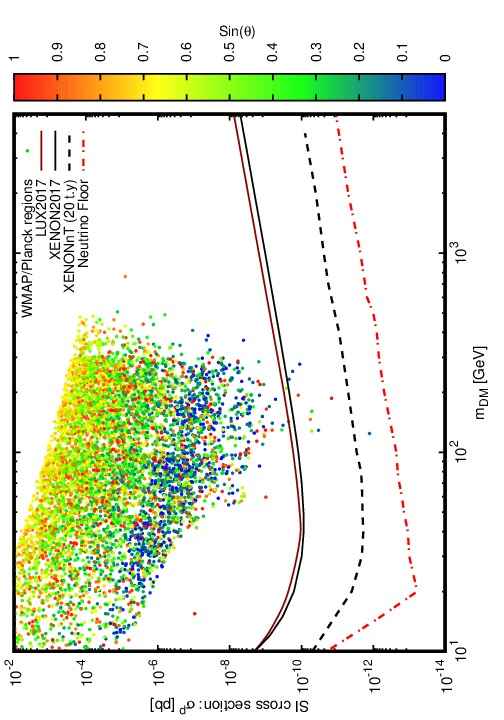}
\end{subfigure}
\begin{subfigure}
\centering
\includegraphics[width=.38\textwidth,angle =-90]{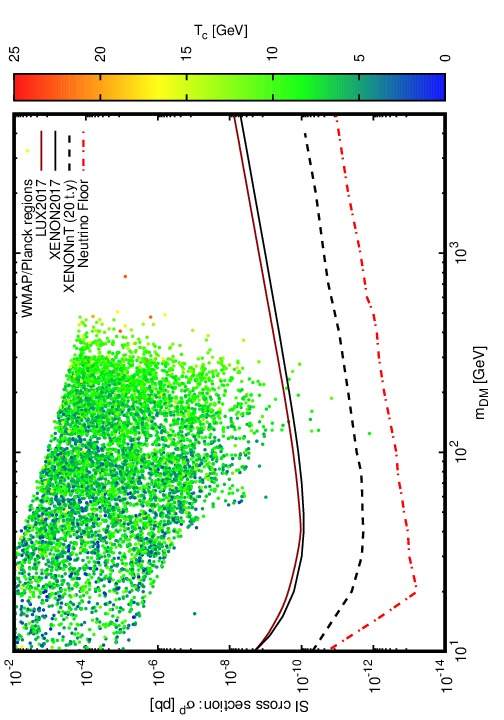}
\end{subfigure}
\caption{Scenario \ref{d1} in model without $s \mbox{-} s'$ cross-coupling terms: In all plots, the spin-independent DM-proton elastic scattering cross section as a function of the DM mass are shown and compared with 
the DD experimental upper limits from LUX, XENON1T, XENONnT projections and
the neutrino background. This phase transition channel gives rise to a DM model which includes only a fraction of the relic density with the DM mass in the range $127-277$ GeV.
The vertical color spectrum in all plots indicates, {\it upper-left)} the fraction of the observed relic abundance, {\it upper-right)} the mass splitting $\delta$, 
{\it lower-left)} the variation of the mixing angle $\sin\theta$ and {\it lower-right)}  the critical temperature $T_{\text{c}}$.}
\label{figd1}
\end{figure}

\begin {table}
{\scriptsize
\begin{tabular}{l*{11}{c}r}

$m_{\phi}$[GeV] & $\delta$[GeV] & $\lambda_{\text{hs}}$ & $\lambda_{\text{hs}'}$ & $\lambda_{\text{s}}$  & $\lambda_{\text{s}'}$  & $\lambda_{\text{hss}'}$ & $\sin \theta$  & $T_c$[GeV] & $v_c/T_c$ & \% $\Omega_{\phi} h^2 $ &  $\sigma^{\text{p}}_{\text{SI}}$ [pb]\\
\hline
\hline
              127 & 86  & 0.49 & 0.08 & 1.92 & 1.45 & 0.85  & 0.51  & 10.6 & 23.2 & 1.35  &  $5.1\times 10^{-11}$\\
              129 & 181 & 0.08 & 0.37 & 0.27 & 1.3  & 0.64  & 0.12  & 7.8 & 31.5& 17.7 & $3.4\times 10^{-11}$   \\
              146 & 17  & 0.1  & 0.07 & 0.09 & 1.86 & 0.16  & 0.84  & 6.6  & 37.2 & 16.6  &  $5.6\times 10^{-11}$ \\
              161 & 111 & 0.26 & 0.40 & 1.98 & 1.75 & 0.92  & 0.31  & 9.3 & 26.5 & 4.5 & $5.2\times 10^{-11}$    \\
              186 & 378 & 0.09 & 0.43 & 0.46 & 1.7  & 2.83  & 0.98  & 11.1 &  22 & 0.27  &  $1.4\times 10^{-11}$ \\
              210 & 53  & 0.32 & 0.23 & 1.9 & 0.72 & 0.73  & 0.5   &  8.9  &  27.6 & 25.1 &   $2.5\times 10^{-10}$  \\
              255 & 420 & 0.2  & 0.36 & 0.8 & 1.64 & 2.16  & 0.08  &  11.5 &  21.3 &  3.18   & $1.7\times 10^{-10}$  \\
              277 & 952 & 0.32 & 0.27 & 1.63 & 1.13 & 5.3   & 0.06  & 10.7 & 22.8 & 0.85  &  $1.1\times 10^{-10}$  \\
\hline

\end{tabular}
}
\caption{Benchmarks for scenario \ref{d1} in model without the $s \mbox{-} s'$ cross-coupling terms.}
\label{tabd1}
\end{table}

\section{Numerical Results}\label{numres}

In this section we impose simultaneously all the dark matter constraints from Sec. \ref{dmcons} and the strongly first-order phase transition from Sec. \ref{first-order}, in our numerical computations.  

In order to compute numerically the DM relic density we apply the package 
{ \tt MicrOMEGAs} \cite{Barducci:2016pcb} which requires the implementation of our model into the program {\tt LanHEP} \cite{Semenov:2014rea}.
The WMAP \cite{Hinshaw:2012aka} and Planck \cite{Ade:2013zuv} 
measurements of the cosmic microwave background (CMB) strongly constrain 
the mean density of cold dark matter (CDM).
The recent Planck result yields $\Omega_{\text{CDM}} h^2 = 0.12\pm 0.001$ 
\cite{Aghanim:2018eyx}. In our analysis we assume that the scalar DM candidate fully or partially saturates the observed relic density such that $\Omega_{\phi} h^2 \lesssim \Omega_{\text{CDM}} h^2$.  

\begin{figure}\label{figa2}
\begin{subfigure}
 \centering
\includegraphics[width=.38\textwidth,angle =-90]{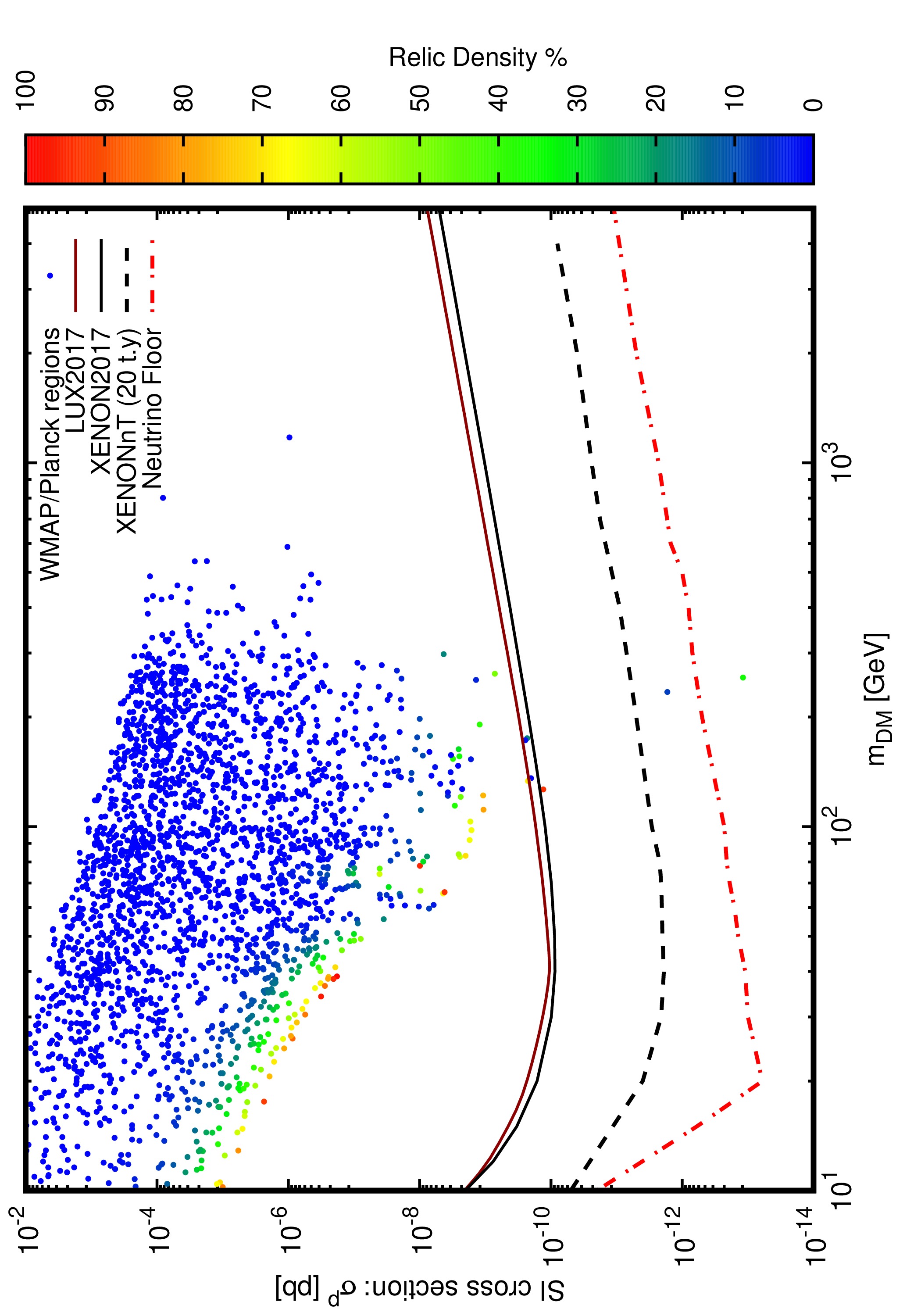}
\end{subfigure}
\begin{subfigure}
\centering
\includegraphics[width=.38\textwidth,angle =-90]{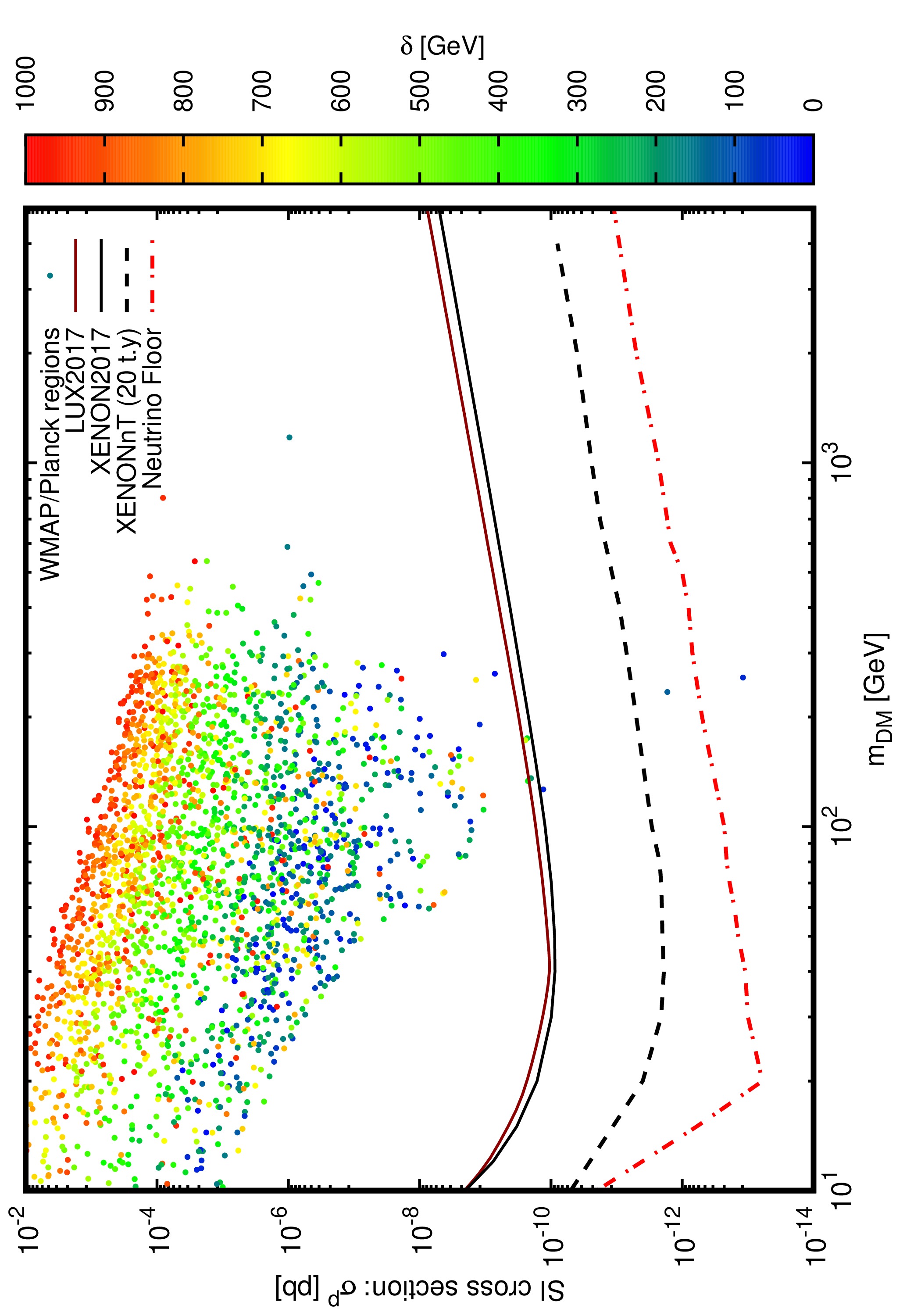}
\end{subfigure}
\newline
\begin{subfigure}
\centering
\includegraphics[width=.38\textwidth,angle =-90]{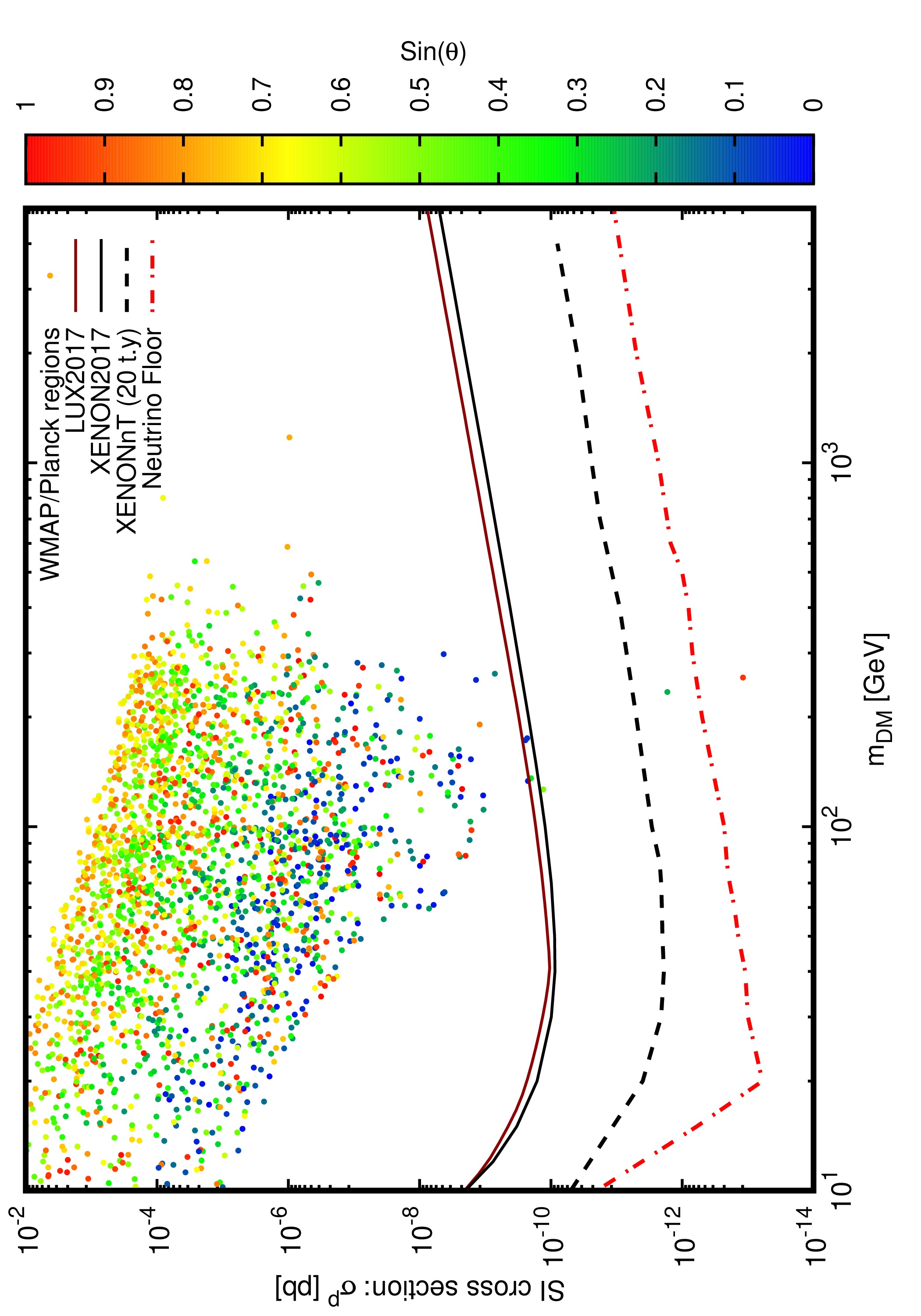}
\end{subfigure}
\begin{subfigure}
\centering
\includegraphics[width=.38\textwidth,angle =-90]{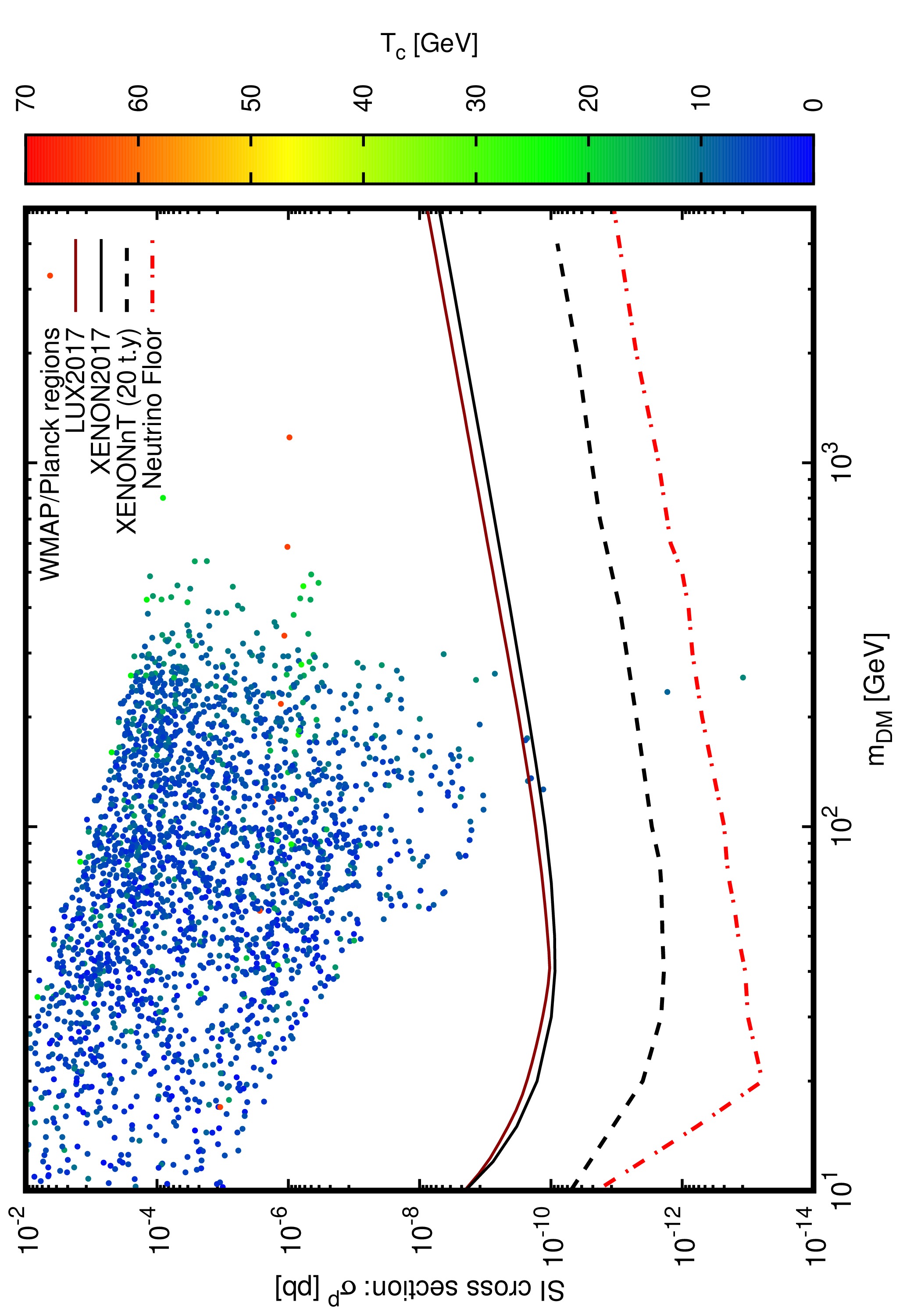}
\end{subfigure}
\caption{Scenario \ref{a2} in model with the $s \mbox{-} s'$ cross-coupling terms: In all plots, the spin-independent DM-proton elastic scattering cross section as a function of the DM mass are shown and compared with 
the DD experimental upper limits from LUX, XENON1T, XENONnT projections and
the neutrino background. This phase transition channel has a narrow viable space respecting all the constraints but remarkably
giving rise to a DM model which consists $75\%$ of the DM relic density.
The vertical color spectrum in the plots indicates, {\it upper-left)} the fraction of the observed relic abundance, {\it upper-right)} the mass splitting $\delta$, 
{\it lower-left)} the variation of the mixing angle $\sin\theta$ and {\it lower-right)}  the critical temperature $T_{\text{c}}$.}\end{figure}

\begin {table}
{\scriptsize
\begin{tabular}{l*{13}{c}r}

$m_{\phi}$[GeV] & $\delta$[GeV] & $\lambda_{\text{hs}}$ & $\lambda_{\text{hs}'}$ & $\lambda_{\text{s}}$  & $\lambda_{\text{s}'}$  & $\lambda_{\text{hss}'}$ & $\lambda_{ss'}$ & $\lambda_{ss'}'$ & $\sin \theta$  & $T_c$[GeV] & $v_c/T_c$ & \% $\Omega_{\phi} h^2 $ &  $\sigma^{\text{p}}_{\text{SI}}$ [pb]\\
\hline
\hline
          126    & 43  & 0.04  & 0.46  & 1.47  & 1.68   &  0.35  & 0.78 & 1.47 & 0.48 &  7.8  &  31.5  & 93.7  &  $1.3\times 10^{-10}$ \\
          234    & 127 & 0.24 & 0.45  & 1.17  & 1.63   & 1.12   & 1.81 & 1.02 & 0.23 &  9.14  &  27  & 8.03  &  $1.7\times 10^{-12}$   \\ 
          256    & 54  & 0.1 & 0.24  &  1.08 &  0.45  &  0.67  & 1.83 & 1.08 & 0.93 &  12.3  &  20  & 35.2  & $1.2\times 10^{-13}$     \\
          
\hline
\end{tabular}
}
\caption{Benchmarks for scenario \ref{a2} in the model with the $s \mbox{-} s'$ cross-coupling terms. }
\label{taba2}
\end{table}

The DM phenomenology of the present model is fully studied in \cite{Ghorbani:2014gka}. However, we recap some main results therein.
We recall that in the simplest extension to the SM, with a singlet scalar 
DM candidate, except the resonance region the rest of the parameter space 
is excluded by the recent direct detection (DD) bounds. 
One of the characteristics that the two-scalar DM model inherits 
and is absent in the single scalar model is manifested by the regions in the parameter space which evade the current DD upper limits. 

In the single scalar model, the DM-nucleon scattering cross section 
and the annihilation cross section are both proportional to a single coupling constant. Regions in the parameter space with large enough 
coupling constant giving rise to the correct relic abundance, have large DD scattering cross section which are excluded by the present DD experiments.

In our extended scalar model, when two particles in the final state, there is a DM annihilation process with 
a heavy WIMP mediated in $t$- or $u$- channel, see the top-left diagram in Fig.~\ref{anni-diagrams}. The presence of this process is critical in the analysis, because this process inters a contribution with a coupling other than 
that in the DD cross section.
Therefore it becomes plausible to find viable regions in the parameter 
space with small coupling for dark matter elastic scattering cross section 
and hence small DD cross section, and 
at the same time large enough dark matter annihilation coupling to induce the correct 
DM relic abundance. 

\begin{figure}
\begin{subfigure}
 \centering
\includegraphics[width=.38\textwidth,angle =-90]{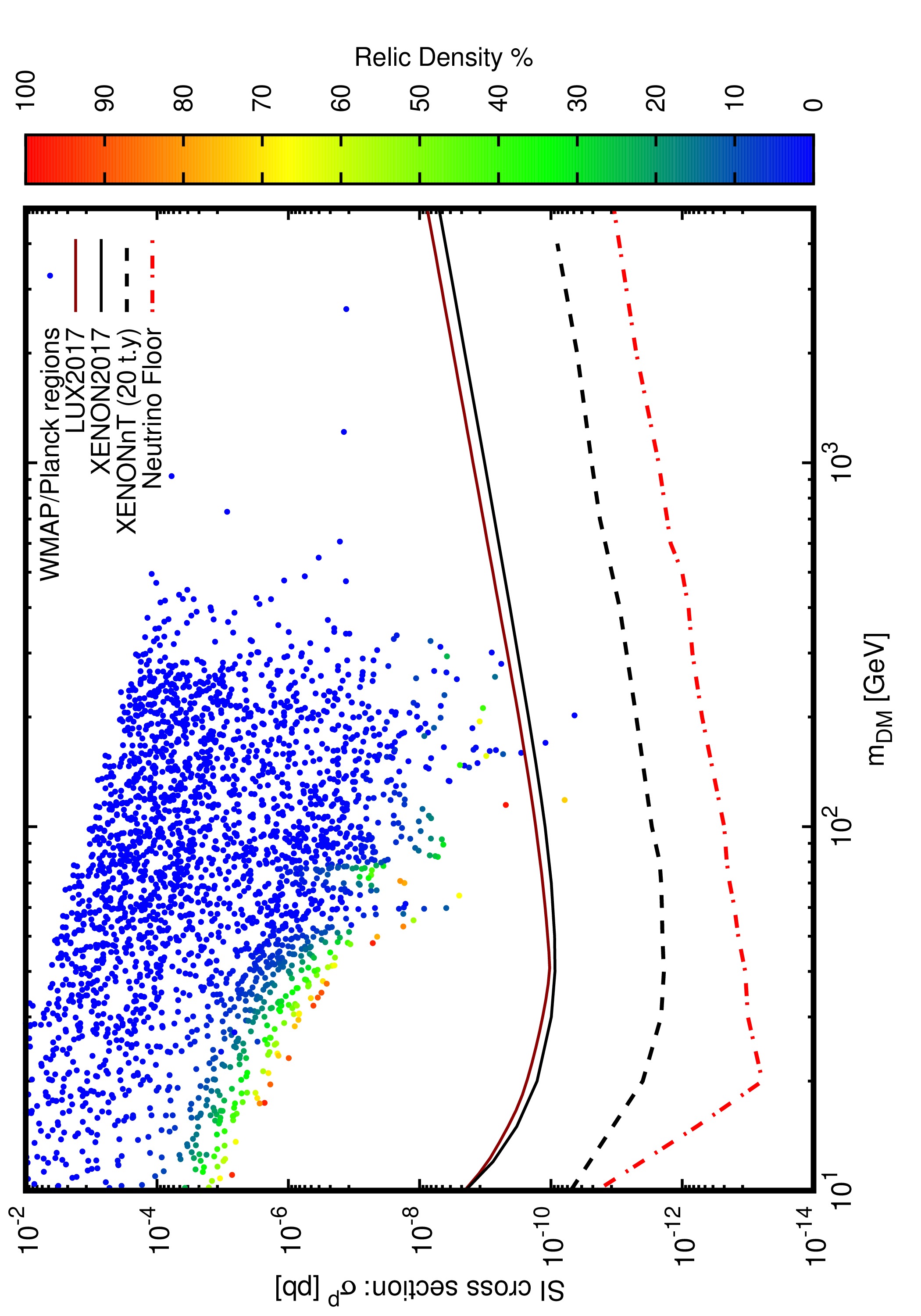}
\end{subfigure}
\begin{subfigure}
\centering
\includegraphics[width=.38\textwidth,angle =-90]{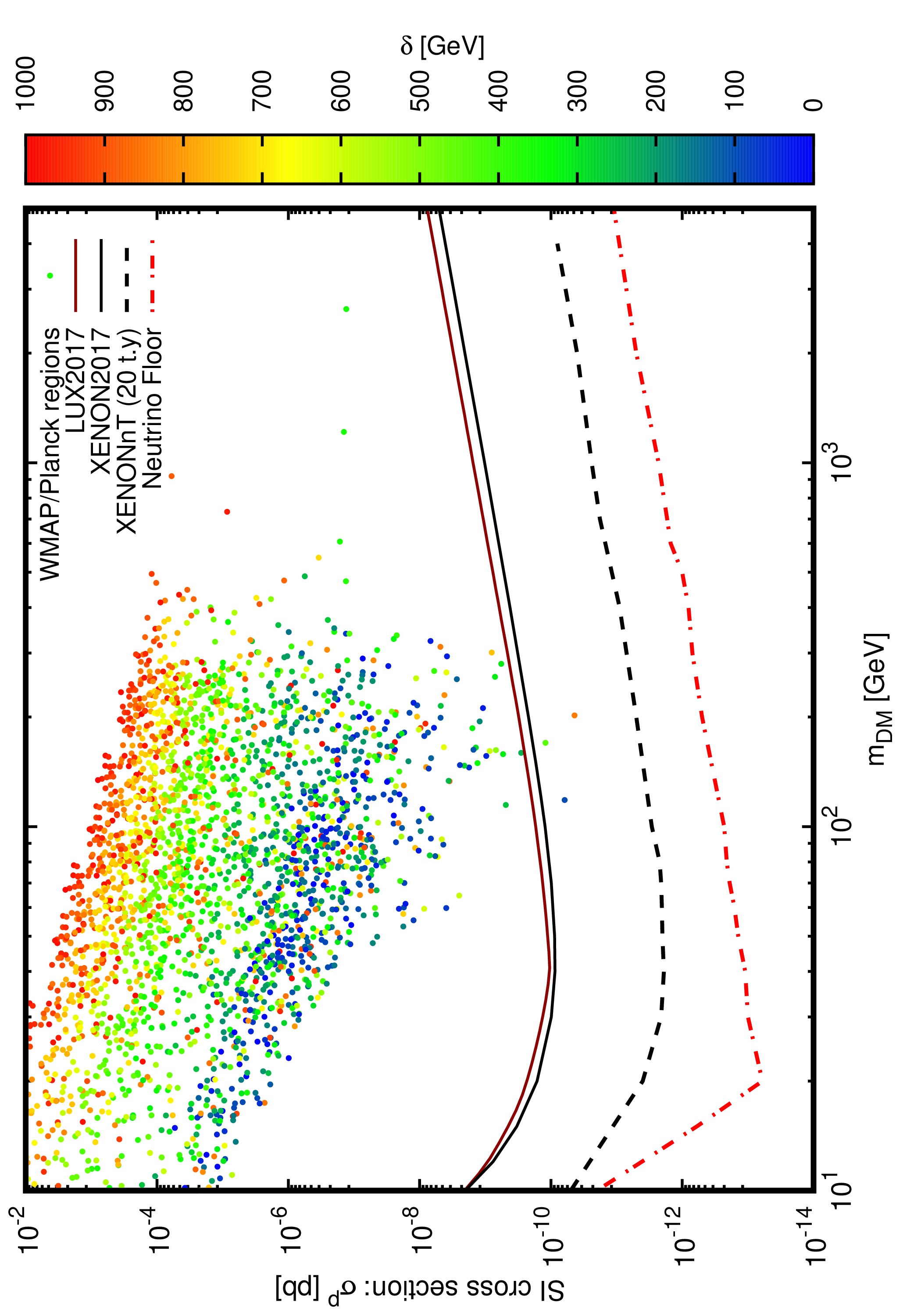}
\end{subfigure}
\newline
\begin{subfigure}
\centering
\includegraphics[width=.38\textwidth,angle =-90]{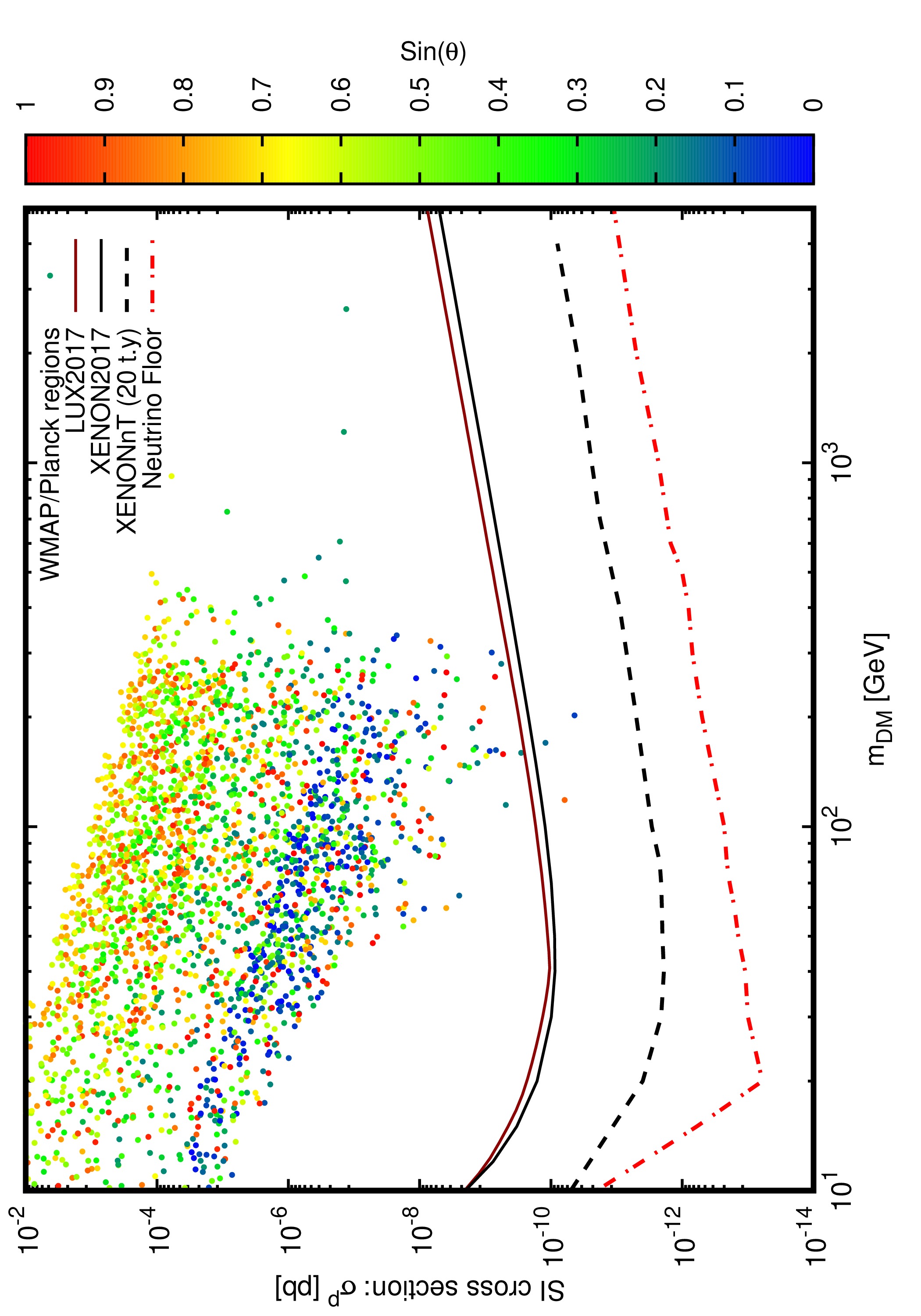}
\end{subfigure}
\begin{subfigure}
\centering
\includegraphics[width=.38\textwidth,angle =-90]{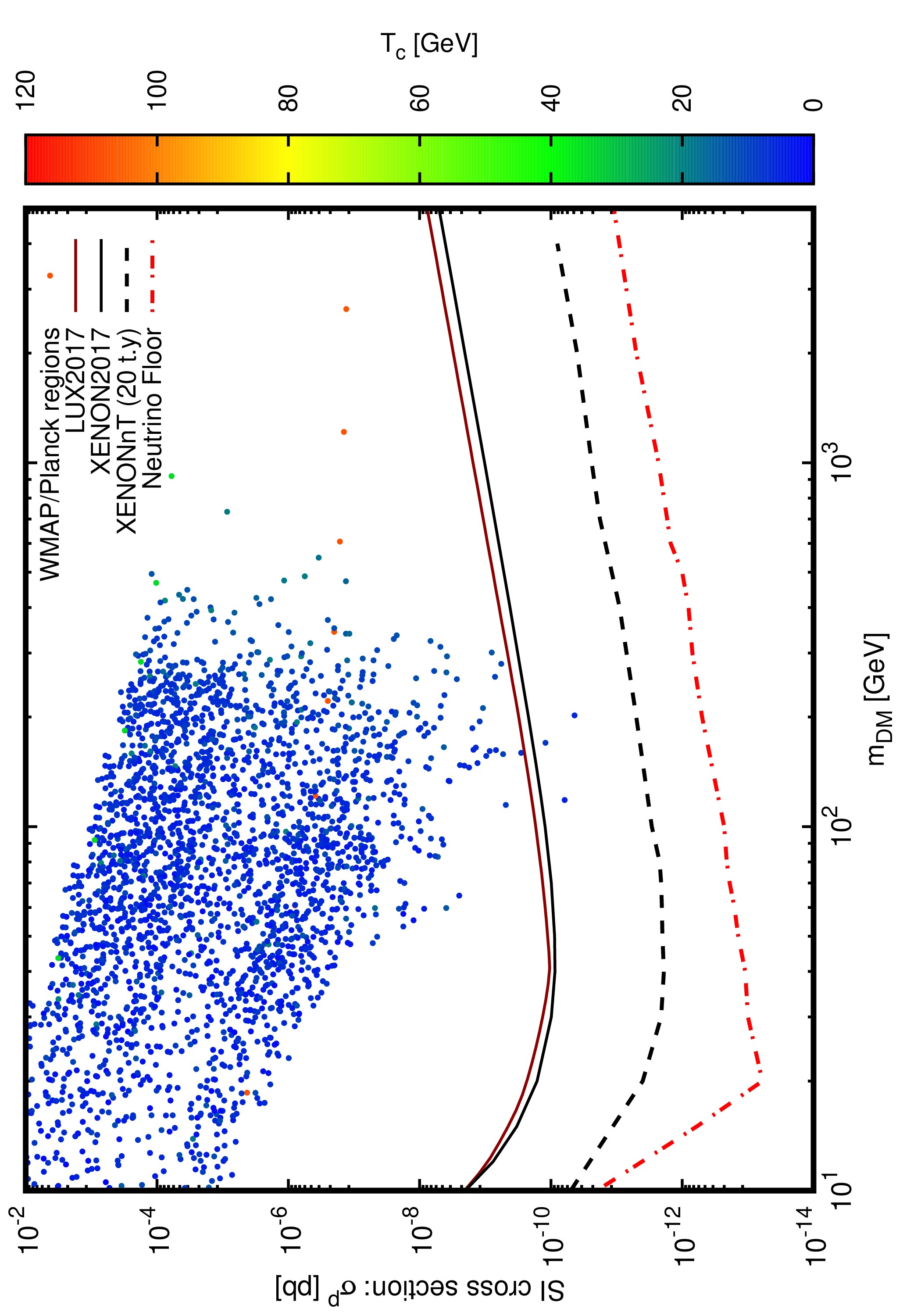}
\end{subfigure}
\caption{Scenario \ref{b2} in model with the $s \mbox{-} s'$ cross-coupling terms: In all plots, the spin-independent DM-proton elastic scattering cross section as a function of the DM mass are shown and compared with 
the DD experimental upper limits from LUX, XENON1T, XENONnT projections and
the neutrino background. This phase transition channel is very similar to the scenario \ref{a2} with a difference that here the DM scalar takes non-zero VEV before the EWPT while in \ref{a2} the DM scalar always has a vanishing VEV.
The vertical color spectrum in all plots indicates, {\it upper-left)} the fraction of the observed relic abundance, {\it upper-right)} the mass splitting $\delta$, 
{\it lower-left)} the variation of the mixing angle $\sin\theta$ and {\it lower-right)}  the critical temperature $T_{\text{c}}$.}
\label{figb2}
\end{figure}

\begin {table}
{\scriptsize
\begin{tabular}{l*{13}{c}r}

$m_{\phi}$[GeV] & $\delta$[GeV] & $\lambda_{\text{hs}}$ & $\lambda_{\text{hs}'}$ & $\lambda_{\text{s}}$  & $\lambda_{\text{s}'}$  & $\lambda_{\text{hss}'}$ & $\lambda_{ss'}$ & $\lambda_{ss'}''$ & $\sin \theta$  & $T_c$[GeV] & $v_c/T_c$ & \% $\Omega_{\phi} h^2 $ &  $\sigma^{\text{p}}_{\text{SI}}$ [pb]\\
\hline
\hline
    118  &  101 & 0.27 & 0.45  & 0.56  & 0.79  & 0.95 & 1.74 & 0.89 & 0.87  & 5.02 & 48.9 & 73.3  & $6.18\times 10^{-11}$ \\
    202  &  839 & 0.46 & 0.44  & 1.63  & 1.46 & 5.58  & 1.69 & 0.82 & 0.08  & 6.63 & 37   &  0.23 & $4.4 \times 10^{-11}$   \\ 
    170  &  444 & 0.44 & 0.29  & 1.52  & 0.72 & 3.15  & 1.6  & 0.28 & 0.14  & 8.24 & 29.8 & 0.24  & $1.2 \times 10^{-10}$     \\
          
\hline
\end{tabular}
}
\caption{Benchmarks for scenario \ref{b2} in the model with the $s \mbox{-} s'$ cross-coupling terms.}
\label{tabb2}
\end{table}

We consider two models in the following analysis. 
In the first case the $s \mbox{-} s'$ cross-coupling terms are absent 
in the Lagrangian, i.e.  $\lambda_{\text{ss}'} = \lambda'_{\text{ss}'} =  \lambda''_{\text{ss}'} = 0$ as discussed in subsection \ref{nnself} to obtain the  first-order EWPT in the model without $s \mbox{-} s'$ cross-coupling terms. The independent free parameters are 
$\lambda_{\text{hs}}$, $\lambda_{\text{hs}'}$, $\lambda_{\text{s}}$, $\lambda_{\text{s}'}$, $m_{\text{s}}$, 
$m_{\text{s}'}$ and the mixing angle $\theta$. 
The coupling constant $\lambda_{\text{hss}'}$ is given in terms of the mixing angle and WIMP masses in Eq. (\ref{tan}). 
In the second scenario studied in subsection \ref{selfint}, the $s \mbox{-} s'$ cross-coupling terms are included and
the dimension of the parameter space is increased. The set of
the free parameters in this case is $\lambda_{\text{hs}}$, $\lambda_{\text{hs}'}$, $\lambda_{s}$, $\lambda_{\text{s}'}$,   $\lambda_{\text{ss}'}$,$\lambda'_{\text{ss}'}$, $\lambda''_{\text{ss}'}$, $m_{\text{s}}$, $m_{\text{s}'}$, and the mixing angle $\theta$.
In all phase transition scenarios discussed in subsections \ref{nnself} and \ref{selfint} 
we perform a full scan with $1.6\times 10^{8}$ samplings 
over the parameter space in the following parameter intervals: 
10 GeV $< m_{\phi} < $ 5 TeV, $m_{\phi'} = m_{\phi}+\delta$, 1 GeV $<\delta<$ 1 TeV, $0 < \lambda_{\text{hs}}, \lambda_{\text{hs}'} < 1$, $0 < \lambda_{\text{s}},\lambda_{\text{s}'}  < 2$, 
$0< \sin~\theta < 1$, and when relevant,
$0< \lambda_{\text{ss}'}, \lambda'_{\text{ss}'}, \lambda''_{\text{ss}'} < 2$.

In the model without the $s \mbox{-} s'$ cross-coupling terms there are four scenarios for the electroweak phase transition. The first scenario \ref{a1}, does not give rise to a first-order phase transition because of an internal inconsistency in the first-order conditions. For scenarios \ref{b1} and \ref{c1}, the first-order phase transition conditions are too restrictive to overlap with that of the dark matter relic density even for a tiny fraction of the DM relic density. Therefore neither a transition from $(0,w,0)$ nor $(0,0,w')$ into $(v,0,0)$ can occur in the  model without the $s \mbox{-} s'$ cross-coupling terms. The last scenario \ref{d1} in this model, i.e. from $(0,w,w')$ to $(v,0,0)$ as seen in Fig. \ref{figd1}, has a viable parameter space. In Fig. \ref{figd1}, all four plots illustrates the viable DM mass against the DM-nucleon cross section with the color spectrum indicating the relic density percentage (upper-left),  the WIMP's mass deference $\delta$ (upper-right), the mixing angle parameter $\sin\theta$ (lower-left) and the critical temperature $T_c$ (lower-right). The upper-left plot shows that the DM mass takes values in the range $~127-277$ GeV (see Table \ref{tabd1}) to evade the direct detection experiments LUX2017/XENON1T, and to be still in the access of the XENONnT and above the neutrino floor. In upper-right plot, the parameter $\delta$ which is the mass deference between the DM scalar, $s$, and the heavy scalar, $s'$, takes a wide range being from a few GeV to around $1$ TeV. Similarly the mixing angle in lower-left plot in Fig. \ref{figd1} takes all values between zero and one. Finally the lower-right plot shows that the critical temperature is of order $10$ GeV. Note that it has been assumed that the phase transition takes place above the DM freeze-out temperature. In Table \ref{tabd1} a list of benchmarks have been represented. The maximum percentage of DM relic density that can be accounted by the scalar $s$, is $\sim 25\%$ for the DM with mass of $\sim 210$ GeV and $T_c\sim 9$ GeV.  It should be noted also that from the ratio $v_c/T_c$ in Table \ref{tabd1}, it is obvious that the phase transition is very strong.

\begin{figure}
\begin{subfigure}
 \centering
\includegraphics[width=.38\textwidth,angle =-90]{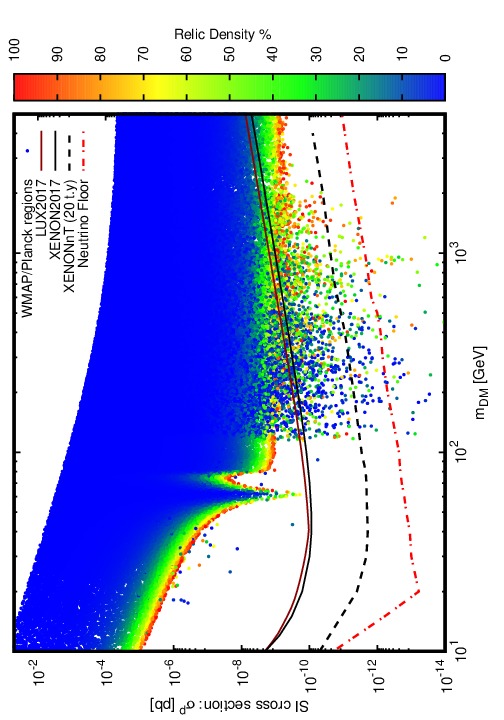}
\end{subfigure}
\begin{subfigure}
\centering
\includegraphics[width=.38\textwidth,angle =-90]{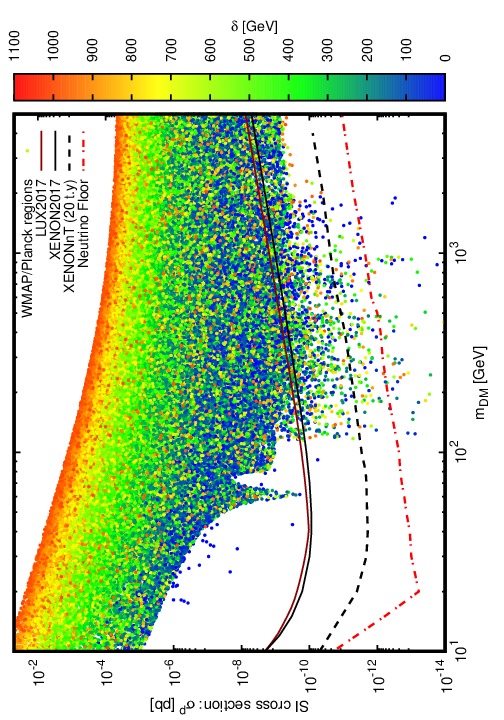}
\end{subfigure}
\newline
\begin{subfigure}
\centering
\includegraphics[width=.38\textwidth,angle =-90]{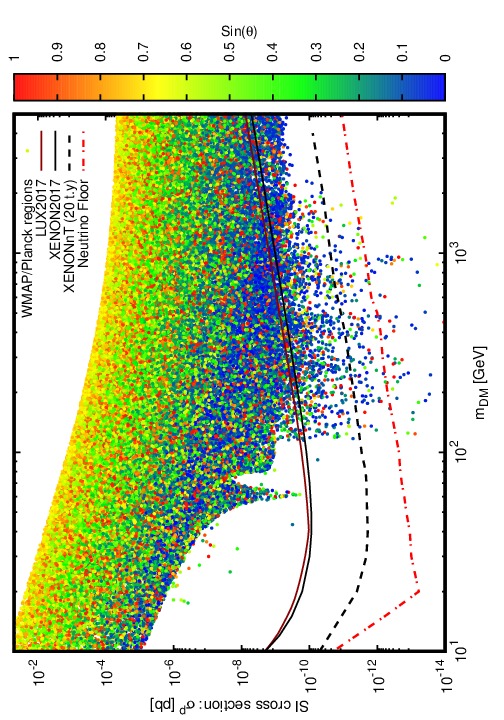}
\end{subfigure}
\begin{subfigure}
\centering
\includegraphics[width=.38\textwidth,angle =-90]{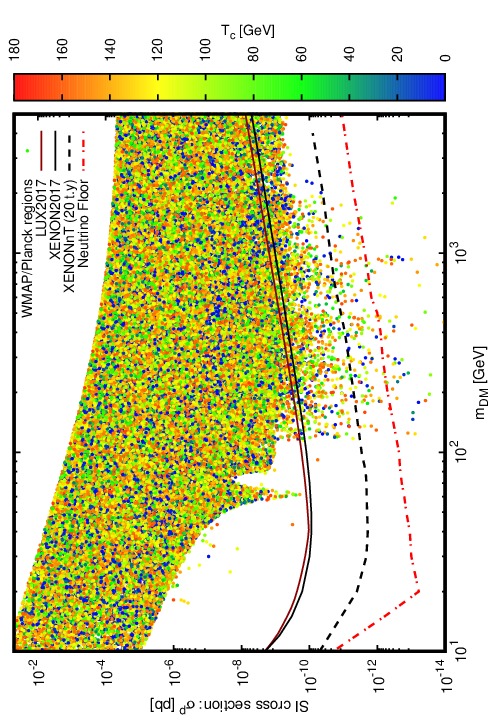}
\end{subfigure}
\caption{Scenario \ref{c2} in the model with the $s \mbox{-} s'$ cross-coupling terms: In all plots, the spin-independent DM-proton elastic scattering cross section as a function of the DM mass are shown and compared with 
the DD experimental upper limits from LUX, XENON1T, XENONnT projections and
the neutrino background. This is a phase transition channel which can explain all the DM content in the universe giving a viable DM mass from  $240 $ GeV to a few TeV.
The vertical color spectrum in all plots indicates, {\it upper-left)} the fraction of the observed relic abundance, {\it upper-right)} the mass splitting $\delta$, 
{\it lower-left)} the variation of the mixing angle $\sin\theta$ and {\it lower-right)}  the critical temperature $T_{\text{c}}$.}
\label{figc2}
\end{figure}

   \begin {table}
{\scriptsize
\begin{tabular}{l*{13}{c}r}

$m_{\phi}$[GeV] & $\delta$[GeV] & $\lambda_{\text{hs}}$ & $\lambda_{\text{hs}'}$ & $\lambda_{\text{s}}$  & $\lambda_{\text{s}'}$  & $\lambda_{\text{hss}'}$ & $\lambda_{ss'}$ & $\sin \theta$  & $T_c$[GeV] & $v_c/T_c$ & \% $\Omega_{\phi} h^2 $ &  $\sigma^{\text{p}}_{\text{SI}}$ [pb]\\
\hline
\hline

243   &  761 & 0.03  & 0.08  & 1.51  & 0.69  & 1.25  & 1.08  & 0.02  &  106.4 & 2.05  & 91.0 & $3.3 \times 10^{-12} $  \\
258   &  67  & 0.13  & 0.11  & 0.38  & 1.73  & 0.59  & 1.07  & 0.23  &  130.9 & 1.52  & 62.9 & $1.0  \times 10^{-12}$\\
124   &  59  & 0.24  & 0.21  & 1.73  & 0.93  & 0.53  & 1.33  & 0.51  &  85.0  & 2.65  & 98.7 & $8.0  \times 10^{-12}$\\
190   &  17  & 0.03  & 0.15  & 1.92  & 0.93  & 0.13  & 1.45  & 0.29  &  89.0  & 2.54  & 83.9 & $3.9  \times 10^{-11}$\\
278   &  47  & 0.22  & 0.11  & 1.68  & 0.42  & 0.61  & 0.84  & 0.34  &  147.1 & 1.24  & 88.0 & $4.4  \times 10^{-11}$\\
412   &  17  & 0.03  & 0.18  & 1.66  & 0.58  & 0.03  & 0.99  & 0.03  &  110.5 & 1.94  & 90.8 & $2.3  \times 10^{-10}$\\
402   &  742 & 0.03  & 0.23  & 1.56  & 0.91  & 1.75  & 1.38  & 0.02  &  99.2  & 2.21  & 93.6 & $7.6  \times 10^{-12}$\\
572   &  553 & 0.09  & 0.20  & 1.64  & 0.65  & 2.23  & 1.33  & 0.04  &  152.4 & 1.17  & 99.7  & $8.7  \times 10^{-12}$\\
863   &  369 & 0.16  & 0.34  & 1.95  & 1.23  & 3.11  & 1.86  & 0.06  &  107.7 & 1.96  & 90.9 & $2.9  \times 10^{-11}$\\
1050  &  11  & 0.05  & 0.30  & 0.99  & 1.11  & 0.06  & 1.12  & 0.04  &  61.0  & 3.87  & 96.7 & $7.0  \times 10^{-11}$\\
2676  &  2   & 0.35  & 0.18  & 1.9   & 0.73  & 0.06  & 1.18  & 0.11  &  152.4 & 1.1  & 95.0 & $5.8  \times 10^{-10}$\\        
\hline
\end{tabular}
}
\caption{Benchmarks for scenario \ref{c2} in the model with the $s \mbox{-} s'$ cross-coupling terms. }
\label{tabc2}
\end{table}

The model with $s \mbox{-} s'$ cross-coupling terms consists of three scenarios that for each one we have found a viable space of parameters. In scenario \ref{a2} i.e. for a phase transition from $(0,0,w')$ to $(v,0,0)$ as it is seen in Fig. \ref{figa2}, the viable DM mass lies in the range $126-256$ GeV. This DM viable mass is comparable with the scenario \ref{d1} in the  model without the $s \mbox{-} s'$ cross-coupling terms, although these are in two different phase transition channels. From the benchmark in Table \ref{taba2} we see that remarkably the scenario includes a point in the narrow viable space of parameters with the DM mass $126$ GeV which covers almost all the DM content of the universe. The second phase transition channel \ref{b2} has almost the same results as the scenario \ref{a2} as seen in Fig. \ref{figb2} and Table \ref{tabb2}. The deference between the two is that in the latter it is the DM scalar, $s$, that undergoes a non-zero VEV before the EWPT while in scenario \ref{a2}, the DM scalar takes zero VEV before and after the EWPT. The maximum percent of the DM relic abundance is given by a DM mass of about $118$ GeV. Another difference between the two scenarios \ref{a2} and \ref{b2} is that the phase transition for \ref{a2} occur in a higher temperature at $\sim 10$ GeV in comparison to \ref{b2} that the critical temperature is in average $\sim 5$ GeV. Again alike \ref{a2}, the phase transition for the scenario \ref{b2} is very strong with $v_c/T_c\sim 40$. The last phase transition channel in the model with the $s \mbox{-} s'$ cross-coupling terms is from $(0,w,w')$ to $(v,0,0)$ studied in \ref{c2}. It is shown in Fig. \ref{figc2} that for this phase transition scenario there is a larger viable space of parameters with respect to scenarios \ref{a2} and \ref{b2}. In Table \ref{tabc2} some benchmarks are presented that show the fact that such transition in fact is able to accommodate all the observed DM content. The plots in Fig. \ref{figc2} demonstrate the DM mass against the DM-nucleon cross section with the color spectrum being the DM relic density (upper-left), the mass deference parameter $\delta$ (upper-right), the mixing angle $\sin\theta$ (lower-left) and the critical temperature, $T_c$, (lower-right). The viable space consists of DM masses from $\sim 240$ GeV to about $2.7$ TeV if the scalar $s$ covers all content of the dark matter, and to more than $4$ TeV if the scalar $s$ takes a fraction of the DM relic density. The critical temperature in this scenario is higher in comparison with  scenarios \ref{a2} and \ref{b2} being of order $\sim 60-150$ GeV.

The benchmarks represented in all the tables give at least a fraction of the DM relic density and at the same time are consistent with a strong first-order phase transition while evading the restrictive direct detection bounds e.g. from LUX2017/XENON1T and survive also from the invisible Higgs decay constraint. Despite the very restrictive constraints from the first-order phase transition and the direct detection bounds, we observe that the two-scalar model predicts models of dark matter that remarkably evades all the constraints simultaneously.

\section{Conclusion}\label{conc}

In this paper an extension to the SM with two real singlet scalar (dubbed as two-scalar scenario) denoted here by $s$ and $s'$ has been investigated to examine whether the model is capable to accommodate simultaneously several constraints from thermal processes such as the relic density of dark matter and the strongly first-order electroweak phase transition in the early universe to constraints from the direct detection experiments and the invisible Higgs decays bound at the LHC. It is known from the literature that the single scalar extension of the SM fails to explain simultaneously the following constraints: the observed relic density, the first-order EWPT and the invisible Higgs decay width limit. However in \cite{Ghorbani:2018yfr} it was shown that two sets of conditions from the DM relic density and the first-order EWPT are not in fact in conflict in the single scalar model but the space parameter shrinks to regions with a few percent of the DM relic density when the invisible Higgs decay constraint is imposed. 
We have shown in two-scalar model that there can be different phase transition channels from the symmetric phase to broken phase of the Higgs vacuum. Despite very restrictive constraints from the first-order EWPT conditions which is more restrictive than the EWPT condition in the single scalar model, some of the channels in the two-scalar model can explain a  fraction or the whole observed DM relic density, and at the same time the strongly first-order EWPT, the direct detection bounds from LUX/XENON1T and the invisible Higgs decay constraint. We have also represented the benchmarks for each phase transition scenario showing the viable range of the DM mass and all the corresponding parameters.

\appendix
\section{Minima in 3-Dimensional VEV Space}\label{minima}
The most general three-level potential we have considered in this paper consists of two extra singlet scalars in addition to the Higgs field. Taking into account the thermal contributions (the one-loop contribution is negligible) we have, 

\begin{equation}\label{pot}
\begin{split}
 V_{\text{eff}}(h,s,s';T)=-\frac{1}{2} \mu^2_\text{h}(T) h^2 + \frac{1}{4} \lambda_\text{h} h^4 \\
 -\frac{1}{2} \mu^2_\text{s}(T) s^2 + \frac{1}{4} \lambda_\text{s} s^4 
 -\frac{1}{2} \mu^2_{\text{s}'}(T) s'^2 + \frac{1}{4} \lambda_{\text{s}'} s'^4 \\
 + \frac{1}{2} \lambda_{\text{hs}} h^2 s^2 +\frac{1}{2} \lambda_{\text{hs}'} h^2 s'^2 
 +\frac{1}{2}\lambda_{\text{hss}'} s s' h^2 \\
 +\frac{1}{2}\lambda_{\text{ss}'} s^2 s'^2 + \frac{1}{3}\lambda'_{\text{ss}'} s s'^3 + \frac{1}{3}\lambda''_{\text{ss}'} s^3 s'\,,
 \end{split}
 \end{equation}
where
\begin{equation}\label{muhss}
 \mu^2_\text{h}(T)=\mu^2_\text{h} - c_\text{h} T^2,~~
 \mu^2_\text{s}(T)=\mu^2_\text{s} - c_\text{s} T^2,~~
 \mu^2_{\text{s}'}(T)=\mu^2_{\text{s}'} - c_{\text{s}'} T^2\,.
\end{equation}
Let us assume that the extremum of this potential is located at $(v,w,w')$, then the first derivatives at this point is vanishing, 
\begin{equation}\label{extermum}
 V' _\text{h} \equiv \frac{\partial V}{\partial h}\Big\vert_{(v,w,w')}=0,~~
 V' _\text{s} \equiv \frac{\partial V}{\partial s}\Big\vert_{(v,w,w')}=0,~~
 V' _{\text{s}'}\equiv \frac{\partial V}{\partial s}\Big\vert_{(v,w,w')}=0\,.
\end{equation}
Eq. (\ref{extermum}) leads to the following set of equations, 
\begin{subequations}\label{eqset}
 \begin{align}
 &v \left(-\mu_\text{h}^2(T) +\lambda_\text{h} v^2 + \lambda_\text{hs} w^2 + \lambda_{\text{hs}'} w'^2 +\lambda_{\text{hss}'} w w'\right) =0 \label{eqset1}\,,\\
 & -\mu_\text{s}^2(T) w +\lambda_\text{s} w^3 + \lambda_\text{hs} w v^2 + \frac{1}{2}\lambda_{\text{hss}'}  w' v^2 + \lambda_{\text{ss}'} w w'^2 +\frac{1}{3} \lambda'_{\text{ss}'} w'^3 + \lambda''_{\text{ss}'} w^2 w'=0 \,,\\
 & -\mu_{\text{s}'}^2(T) w' +\lambda_{\text{s}'} w'^3 + \lambda_{\text{hs}'} w' v^2 + \frac{1}{2}\lambda_{\text{hss}'}  w v^2 + \lambda_{\text{ss}'} w' w^2 + \lambda'_{\text{ss}'} w w'^2 + \frac{1}{3} \lambda''_{\text{ss}'} w^3 =0\,.
 \end{align}
\end{subequations}
As seen from Eq. (\ref{eqset1}), both $v=0$ and $v\neq 0$ are extrema of the potential. 
Let us also define the second derivatives of the potential at the extremum point $(v,w,w')$ as the following, 
\begin{subequations}\label{secder}
 \begin{align}
&V''_\text{hh}\equiv \frac{\partial^2 V}{\partial h^2} \Big\vert_{(v,w,w')} = -\mu_\text{h}^2(T) +3\lambda_\text{h} v^2 + \lambda_\text{hs} w^2 + \lambda_{\text{hs}'} w'^2 +\lambda_{\text{hss}'} w w' \,,\\
&V''_\text{ss}\equiv \frac{\partial^2 V}{\partial s^2} \Big\vert_{(v,w,w')} = -\mu_\text{s}^2(T) +3\lambda_\text{s} w^2 + \lambda_\text{hs} v^2 + \lambda_{\text{ss}'} w'^2 + \lambda''_{\text{ss}'} w^2 \,,\\
&V''_{\text{s}'\text{s}'}\equiv \frac{\partial^2 V}{\partial s'^2} \Big\vert_{(v,w,w')} = -\mu_{\text{s}'}^2(T) +3\lambda_{\text{s}'} w'^2 + \lambda_{\text{hs}'} v^2 + \lambda_{\text{ss}'} w^2 + 2 \lambda'_{\text{ss}'} w w' \,, \\
&V''_\text{hs}\equiv \frac{\partial^2 V}{\partial h \partial s} \Big\vert_{(v,w,w')} = 2 \lambda_\text{hs} vw+\lambda_{\text{hss}'} vw' \,, \\
&V''_{\text{hs}'}\equiv \frac{\partial^2 V}{\partial h \partial s'} \Big\vert_{(v,w,w')} = 2 \lambda_{\text{hs}'} vw'+\lambda_{\text{hss}'} vw \,,\\
&V''_{\text{ss}'}\equiv \frac{\partial^2 V}{\partial s \partial s'} \Big\vert_{(v,w,w')} = \frac{1}{2}\lambda_{\text{hss}'} v^2+2 \lambda_{\text{ss}'} ww'+\lambda'_{\text{ss}'} w'^2+\lambda''_{\text{ss}'} w^2\,.
 \end{align}
\end{subequations}

The conditions for the point $(v,w,w')$ to be a local minimum are, 
\begin{equation}\label{mincon}
 V''_\text{hh}>0,~~
 \begin{vmatrix}
V''_\text{hh} &  V''_\text{hs} \\ 
 V''_\text{hs} &  V''_\text{ss} \\ 
\end{vmatrix} > 0\,,~
\begin{vmatrix}
V''_\text{hh} & V''_\text{hs} & V''_{\text{hs}'} \\ 
V''_\text{hs} & V''_\text{ss} & V''_{\text{ss}'} \\ 
V''_{\text{hs}'} & V''_{\text{ss}'} & V''_{\text{s}'\text{s}'}
\end{vmatrix} >0 \,.
\end{equation}
\subsection{Model without $s \mbox{-} s'$ cross-coupling terms}\label{nnselfcase}
The first case we have considered in this paper is when there is no $s \mbox{-} s'$ cross-coupling terms in the potential in Eq. (\ref{pot}), i.e., the case $\lambda_{\text{ss}'}=\lambda'_{\text{ss}'}=\lambda''_{\text{ss}'}=0$. Eqs. (\ref{eqset}) then is simplified as,
\begin{subequations}\label{eqsetcase1}
 \begin{align}
 v \left(-\mu_\text{h}^2(T) +\lambda_\text{h} v^2 + \lambda_\text{hs} w^2 + \lambda_{\text{hs}'} w'^2 +\lambda_{\text{hss}'} w w'\right) =0 \label{eqset1case1} \,, \\
 -\mu_\text{s}^2(T) w +\lambda_\text{s} w^3 + \lambda_\text{hs} w v^2 + \frac{1}{2}\lambda_{\text{hss}'}  w' v^2 + \lambda_{\text{ss}'} w w'^2 =0 \,, \\
 -\mu_{\text{s}'}^2(T) w' +\lambda_{\text{s}'} w'^3 + \lambda_{\text{hs}'} w' v^2 + \frac{1}{2}\lambda_{\text{hss}'}  w v^2 + \lambda_{\text{ss}'} w' w^2 =0 \,.
 \end{align}
\end{subequations}
We divide the solutions in Eqs. (\ref{eqsetcase1}) to two classes; the  electroweak  symmetric phase at which the Higgs vacuum expectation value is vanishing, $v=0$, and the broken phase that $v\neq 0$. The solutions in $v=0$ class are obtained as, 

\begin{subequations}
 \begin{align}
&(0,0,0) \,, \\
&(0,0, w'^2=\frac{\mu^2_{\text{s}'}(T)}{\lambda_{\text{s}'}}) \,, \\
&(0, w^2=\frac{\mu^2_{\text{s}}(T)}{\lambda_{\text{s}}},0) \,, \\
&(0, w^2=\frac{\mu^2_{\text{s}}(T)}{\lambda_s },w'^2=\frac{\mu^2_{\text{s}'}(T)}{ \lambda_{\text{s}'}}) \,.
 \end{align}
\end{subequations}
In $v\neq 0$ class if $w=0$ then we must have also $w'=0$, and vice verse. Therefore the solutions in this class are only in the following forms, 
\begin{subequations}
 \begin{align}
&(v^2=\frac{\mu^2_\text{h}(T)}{\lambda_\text{h}},0,0) \,, \\
&(v\neq 0, w\neq 0, w'\neq 0) \,.
 \end{align}
\end{subequations}
After the electroweak symmetry breaking the Higgs vacuum expectation value is non-zero, but if the scalar $s$ wants to be the DM candidate it must take zero VEV after the EWPT (or to be more accurate after the DM freeze-out). Therefore, the only vacuum structure of the two-scalar model after the EWPT is $(v^2=\frac{\mu^2_\text{h}(T)}{\lambda_\text{h}},0,0)$.

The extremum $(v,w,w')$ must be also local minimum, at least in some temperature intervals, as has been discussed throughout the paper. The second derivatives at the extremum point $(v,w,w')$ for the model without the $s \mbox{-} s'$ cross-coupling terms read, 
\begin{subequations}\label{secder1}
 \begin{align}
&V''_\text{hh} = -\mu_\text{h}^2(T) +3\lambda_\text{h} v^2 + \lambda_\text{hs} w^2 + \lambda_{\text{hs}'} w'^2 +\lambda_{\text{hss}'} w w' \,, \\
&V''_\text{ss} = -\mu_\text{s}^2(T) +3\lambda_\text{s} w^2 + \lambda_\text{hs} v^2  \,, \\
&V''_{\text{s}'\text{s}'} = -\mu_{\text{s}'}^2(T) +3\lambda_{\text{s}'} w'^2 + \lambda_{\text{hs}'} v^2  \,,\\
&V''_\text{hs} = 2 \lambda_\text{hs} vw+\lambda_{\text{hss}'} vw' \,, \\
&V''_{\text{hs}'} = 2 \lambda_{\text{hs}'} vw'+\lambda_{\text{hss}'} vw' \,,\\
&V''_{\text{ss}'} = \frac{1}{2}\lambda_{\text{hss}'} v^2\,.
 \end{align}
\end{subequations}

\subsection{Model with $s \mbox{-} s'$ cross-coupling terms}\label{selfcase}
In this case, at least one of the $s \mbox{-} s'$ cross-couplings are non-vanishing, i.e. $\lambda_{\text{ss}'}\neq 0$, or $\lambda'_{\text{ss}'}\neq 0$, or $ \lambda''_{\text{ss}'}\neq 0$. The generic VEV set $(v,w,w')$ with $v$ , $w$ and $w'$ being the VEV of the scalar fields, $h$, $s$ and $s'$ respectively, is the extremum of the general potential in Eq. (\ref{pot}) if it satisfies Eq. (\ref{eqset}). Finding all solutions for Eq. (\ref{eqset}) in general is complicated. We therefore study only the simpler solutions some of which are considered also in the model without the $s \mbox{-} s'$ cross-coupling terms, \\

\underline{$\lambda_{\text{ss}'}\neq 0 \,,  \lambda'_{\text{ss}'}=\lambda''_{\text{ss}'}=0 $}
\begin{subequations}\label{ext1nonself}
\begin{align}
& (v=0, w=0, w'^2= \frac{\mu^2_{\text{s}'}(T)}{\lambda_{\text{s}'}} )\,,\\
& (v=0, w^2=\frac{\mu^2_{\text{s}}(T)}{\lambda_{\text{s}}}, w'=0 )\,,\\
& ( v=0, w^2= \frac{\lambda_{\text{s}'} \mu^2_{\text{s}}(T)-\lambda_{\text{ss}'} \mu^2_{\text{s}'}(T)}{\lambda_{\text{s}}\lambda_{\text{s}'}-\lambda_{\text{ss}'}^2}, w'^2= \frac{\lambda_{\text{s}} \mu^2_{\text{s}'}(T)-\lambda_{\text{ss}'} \mu^2_{\text{s}}(T)}{\lambda_{\text{s}}\lambda_{\text{s}'}-\lambda_{\text{ss}'}^2} )\,.
\end{align}
\end{subequations}

\underline{$\lambda_{\text{ss}'}\neq 0\,, \lambda''_{\text{ss}'}\neq 0 \,,\lambda'_{\text{ss}'}=0 $}
\begin{equation}\label{ext2nonself}
(v=0, w=0, w'^2= \frac{\mu^2_{\text{s}'}(T)}{\lambda_{\text{s}'}} )\,.
\end{equation}

\underline{$\lambda_{\text{ss}'}\neq 0\,,\lambda'_{\text{ss}'}\neq 0  \,,\lambda''_{\text{ss}'}=0 $}
\begin{equation}\label{ext3nonself}
(v=0, w^2=\frac{\mu^2_{\text{s}}(T)}{\lambda_{\text{s}}}, w'= 0 )\,.
\end{equation}

After the phase transition that $v\neq 0$ and $w=0$, the only solution to Eq. (\ref{eqset}), similar to the non-interacting case is, 
\begin{equation}\label{ext4nonself}
 (v^2=\frac{\mu^2_{\text{h}}(T)}{\lambda_{\text{h}}},w=0,w'=0)\,.
\end{equation}
All the extremum solutions in Eqs. (\ref{ext1nonself})-(\ref{ext4nonself}) must satisfy the local minimum conditions in Eq. (\ref{mincon}).
\section{Critical Temperature and Deepest Minimum Condition}\label{crt}

Let us represent the VEVs of the scalars before the phase transition i.e. in the symmetric phase, as $(v_\text{sym},w_1,w'_1)$ and after the phase transition i.e. in the broken phase as $(v_\text{brk},w_2,w'_2)$. Note by the symmetric and broken phase we mean only in the electroweak symmetry group $SU(2)$ and we do not in general consider the symmetry status of other scalar field in the theory. 
The critical temperature is defined as the temperature at which the symmetric and broken minima of the thermal effective potential become degenerate, therefore, 

\begin{equation}\label{crtcon}
 V_\text{eff}(\text{sym},w_1,w'_1;T_c)=V_\text{eff}(v_\text{brk},w_2,w'_2;T_c)\,.
\end{equation}

In order for the local minimum $(v_\text{sym},w_2,w'_2)$ to be also the global one, it must be deeper than the local minimum $(v_\text{sym},w_1,w'_1)$, i.e., 
\begin{equation}\label{deltaV}
 \Delta V_\text{eff}(T)\equiv V_\text{eff}(0,w_1,w'_1;T)- V_\text{eff}(v_\text{brk},w_2,w'_2;T)>0 \,,
\end{equation}
which must hold for all $T<T_c$. To require the condition (\ref{deltaV}) to satisfy for $T<T_c$ it is enough that the $T$-derivative of $\Delta V_\text{eff}(T)$ be negative at $T_c$.

\begin{acknowledgments}
PHG was supported in part by AP grant AP-CCR-2019137. Arak University is acknowledged for financial support under contract no.98/664.
\end{acknowledgments}

\bibliographystyle{JHEP}
\bibliography{ref.bib}

\end{document}